\documentclass[12pt,preprint]{aastex}

\usepackage{color}

\newcommand\simgreater{\,\lower0.7ex\hbox{$\stackrel{>}{\sim}$}\,}
\newcommand\simless{\,\lower0.7ex\hbox{$\stackrel{<}{\sim}$}\,}

\newcommand{\gcc}{\ \mathrm{g\ cm^{-3} }}

\newcommand{\microm}{\ \mathrm{\mu m}}

\newcommand{\cmss}{\ \mathrm{cm \ s^{-2}}}

\newcommand{\wpcm}{\ \mathrm{W \ cm^{-2} }}

\slugcomment{\apjs\/ in press}

\shorttitle{On Validating FLASH}
\shortauthors{Calder et~al.}

\begin{document}

\title{On Validating an Astrophysical Simulation Code} 

\author{
A. C. Calder\altaffilmark{1,2},
B. Fryxell\altaffilmark{1,3}, 
T. Plewa\altaffilmark{1,2,4}, 
R. Rosner\altaffilmark{1,2,3}, 
L. J. Dursi\altaffilmark{1,2},
V. G. Weirs\altaffilmark{1,2}, \\ 
T. Dupont\altaffilmark{1,5}, 
H. F. Robey\altaffilmark{6},
J. O. Kane\altaffilmark{6},  
B. A. Remington\altaffilmark{6}, 
R. P. Drake\altaffilmark{7},
G. Dimonte\altaffilmark{6}, \\
M. Zingale\altaffilmark{1,8},
F. X. Timmes\altaffilmark{1,2},
K. Olson\altaffilmark{1,8},
P. Ricker\altaffilmark{1,2}, 
P. MacNeice\altaffilmark{8},
and H. M. Tufo\altaffilmark{1,5} 
}

\altaffiltext{1}{Center for Astrophysical Thermonuclear Flashes, 
                 The University of Chicago, 
                 Chicago, IL  60637}
\altaffiltext{2}{Department of Astronomy \& Astrophysics, 
                 The University of Chicago, 
                 Chicago, IL  60637}
\altaffiltext{3}{Enrico Fermi Institute, 
                 The University of Chicago,
                 Chicago, IL  60637}
\altaffiltext{4}{Nicolaus Copernicus Astronomical Center, 
                 Bartycka 18, 
                 00716 Warsaw, Poland}
\altaffiltext{5}{Department of Computer Science, 
                 The University of Chicago,
                 Chicago, IL  60637}
\altaffiltext{6}{Lawrence Livermore National Laboratory, 
                 Livermore, CA  94550}
\altaffiltext{7}{University of Michigan
                 Ann Arbor, MI  48105}
\altaffiltext{8}{Dept of Astronomy and Astrophysics, The University of 
                 California, Santa Cruz, 
                 Santa Cruz, CA 95064}
\altaffiltext{9}{UMBC/GEST Center, NASA/GSFC,
                 Greenbelt, MD  20771}

\begin{abstract}

We present a case study of validating an astrophysical simulation code.
Our study focuses on validating FLASH, a parallel, adaptive-mesh 
hydrodynamics code for studying the compressible, reactive flows 
found in many astrophysical environments. We describe the 
astrophysics problems of interest and the challenges associated with 
simulating these problems. 
We describe methodology and discuss solutions to 
difficulties encountered in verification and validation.
We describe verification tests regularly administered to the code, 
present the results of new verification tests, 
and outline a method for testing general equations 
of state. We present the results of two validation tests in
which we compared simulations to experimental data.
The first is of a laser-driven shock propagating 
through a multi-layer target, a configuration subject to both Rayleigh-Taylor 
and Richtmyer-Meshkov instabilities. The second test is a classic Rayleigh-Taylor 
instability, where a heavy fluid is supported against the force of gravity by 
a light fluid. 
Our simulations of the multi-layer target experiments showed good
agreement with the experimental results, but our simulations of the Rayleigh-Taylor
instability did not agree well with the experimental results. We discuss 
our findings and present results of additional simulations undertaken to
further investigate the Rayleigh-Taylor instability.

\end{abstract}


\section{Introduction}

The enormous progress seen in the evolution of fast computing machines
and numerical methods stands as one of the great achievements of the
twentieth century. Numerical modeling is now an
accepted and widely applied method of research, and in many cases
simulations have matured to the extent that they now provide
direction to theoretical research. In astrophysics, where the complexity of
many of the problems requires large-scale computing for any hope of progress, much
of research involves the development and application of reliable
and trustworthy simulation codes. Before the scientific community
can have confidence in results from such a code, it must be
subjected to a wide variety of verification and validation tests.
This paper discusses progress in verification and validation of FLASH,
a parallel, adaptive-mesh simulation code for the compressible, reactive flows 
found in many astrophysical environments.

The goal motivating the development of FLASH is to advance the solution
of several astrophysical problems related to thermonuclear flashes on the
surfaces and in the interiors of compact objects. In particular, 
the problems of interest are 
type I X-ray bursts, classical novae, and Type Ia supernovae.
These events all involve the accretion of
material from a companion star onto the surface of the compact star, followed
by the ignition of either the core of the compact star or the material
accreted onto the surface.
The global physical phenomena common to all three of these events include
an accretion flow onto the surfaces of compact stars,
shear flow and Rayleigh-Taylor instabilities~\citep{taylor50,chandra81}
on the stellar surfaces and in the core, ignition of thermonuclear burning
in degenerate matter, development of convection, propagation of nuclear
burning fronts, and expansion of the stellar envelope. An 
understanding of these global phenomena
requires knowledge of the fundamental physical processes involved in each.
Accordingly, much of our scientific effort focuses on research into the
basic ``microphysics." These fundamental processes include turbulence at
large Reynolds and Rayleigh
numbers, fluid instabilities and mixing, convection and the
convective penetration of stable matter at very high densities, thermodynamics in
relativistic and degenerate regimes, the propagation of both subsonic and supersonic
burning fronts, and radiation hydrodynamics.

Verification and validation are fundamental steps in developing
any new technology, whether it be a simulation code like FLASH or
an instrument for observation. 
For simulation technology, the goal of these testing steps is 
assessing the credibility of modeling and simulation.
Considerable work on verification and validation of
simulations has been done in the field of computational fluid
dynamics (CFD), and in the CFD literature  
the terms verification and validation 
have precise, technical meanings~\citep{aiaa98,roache98a,roache98b}.
Verification is taken to mean demonstrating that a 
code or simulation accurately represents the conceptual model.
Validation of a simulation means 
demonstrating that the simulation appropriately describes nature.
The scope of validation is therefore much
larger than that of verification and includes 
comparison of numerical results with experimental or observational data.
In astrophysics, where it is difficult to obtain observations suitable
for comparison to numerical simulations, this process 
can present unique challenges.

In this paper, we describe our efforts at verifying and validating
the hydrodynamics module in FLASH.
We begin by describing the astrophysical problems of interest and
the importance of fluid instabilities in the problems. 
A discussion of the terminology and methodology for
verification and validation (V\&V) follows this description. 
We follow that with a discussion of the challenges
found in verifying and validating astrophysical simulations, and in particular,
the aspects of astrophysical modeling for which it is difficult to apply 
recommended CFD V\&V techniques. In the next section, we present 
verification tests and include an outline of a procedure for testing an 
arbitrary equation of state in the context of numerical hydrodynamics 
schemes. The following section contains a comparison of the results from 
two laboratory experiments with those obtained from simulations, 
and the final section contains discussion and conclusions.

\subsection{Overview of Astrophysical Thermonuclear Flashes} 

Thermonuclear flashes, events of rapid or explosive thermonuclear
burning, occur in a variety of stellar settings. These events include
type I X-ray bursts, classical novae, and Type Ia supernovae, all of 
which involve a close binary system in which matter from a companion star
accretes onto the surface of a compact star (neutron star or white
dwarf).  Either the core of the compact object or the accreted layer
on the surface of the compact object ignites under electron-degenerate
conditions, and a thermonuclear burning front is born and begins to
propagate.

These events provide not only fantastic observational displays, but also 
tools with which potentially to answer several fundamental questions.
The light curves and spectra of X-ray bursts can provide
information about the masses and radii of neutron stars~\citep{lewin93,lamb01} 
and thus also provide information about the nuclear
equation of state. Classical novae can provide information about the
abundances of intermediate-mass elements in the universe and the
dynamics of white dwarfs in close binary systems~\citep{gehrz98}. Type Ia
supernovae provide additional information about the abundances of
intermediate-mass and heavy elements and play a crucial role as
``standard candles" in determining cosmological parameters such as the
Hubble constant, $H_0$, the mass density, $\Omega_M$, and the cosmological
constant or vacuum energy density, $\Omega_\Lambda$~\citep[see][and references 
therein]{riess98,perlmutter98,turner01}.

X-ray bursts are flashes that start at the bottom of a very thin layer 
($\sim 10-100$ m) of hydrogen-rich or helium-rich fuel that has accreted 
onto the surface of a neutron star \citep{taam85,lewin93,taam93}. 
The total energy released by burning the fuel into ash is a
factor of $\sim 20-100$ less than the gravitational binding energy.
Consequently, the accreted material is gravitationally bound to the
neutron star and the flash is not quenched by expansion of the
envelope~\citep{hansen75}.  Instead, fuel in the accreted envelope is incinerated to
iron-peak or heavier nuclei~\citep{shatz01}.

Novae result from the ignition of a layer ($\sim 10^4$ m) of 
hydrogen-rich material that has accreted onto the surface of a white 
dwarf \citep{truran82,shara89,starrfield89,livio94}.  In this case, the total energy
released by thermonuclear burning is a factor of $\sim$ 100 more than
the gravitational binding energy.  As a result, the excess energy
leads to an enormous expansion of the white dwarf's envelope,
which engulfs the companion star and forms a common envelope binary.
The work done against gravity in the expansion of the envelope cools
the hydrogen burning layer and quenches the runaway, leading to the
establishment of a phase of stable hydrogen burning that continues
through envelope exhaustion.

Type Ia supernovae are thought to be due to carbon flashes that ignite
in the cores of accreting white dwarfs \citep{woosley86, nomoto94,
niemeyer95a, niemeyer95b}.  Models involving either a pure
deflagration or a pure detonation have been unable to provide a 
consistent explanation for the observed expansion velocities and the 
spectrum of intermediate-mass and iron-peak ejecta.
Some Type Ia supernova models 
that involve a transition from a deflagration to a detonation
have been constructed.  One possibility is a
more-or-less spontaneous transformation of the initial deflagration
into a detonation as the burning front propagates outward through the
white dwarf \citep{niemeyer95a,niemeyer95b,khokhlov97,niemeyer97}.
Another possibility is that the initial deflagration dies out as a
result of the expansion of the outer layers of the white dwarf.  
When these gravitationally bound layers collapse back onto the white dwarf,
a detonation is ignited \citep{blinnikov87, boisseau96, khokhlov95,
khokhlov97}.  In either case, these models are capable of accounting
for the observed expansion velocities of the silicon-group and
iron-group nuclei.

In all three of these thermonuclear flash events, 
the nuclear burning time scale is much
shorter than the time scale over which the nuclear fuel accretes onto the
surface. These short burning time scales make it likely that 
ignition of the fuel occurs at
either a single point or, at most, at a few discrete points.  The situation
may be complicated by the presence of magnetic fields.  The strong
magnetic fields (B $\sim 10^6$ - $10^9$ G) of white dwarfs and the
super-strong magnetic fields (B $\sim 10^9$ - $10^{12}$ G) of neutron
stars may be capable of funneling the flow of accreting matter onto
the magnetic polar caps of the compact object \citep{lamb73}. The 
accreted matter, which constitutes the nuclear fuel, may or may not be 
able to spread over the surface of the star before ignition occurs.  The
three-dimensional nature of this fuel geometry is compounded by the
effects of a magnetic field on the thermal and mass transport
coefficients \citep{lamb96,potekhin99a,potekhin99b}.

Even in the absence of a magnetic field, thermonuclear flashes are
inherently multi-dimensional because of the complexity of the
underlying fluid instabilities. Most efforts at modeling the
multi-dimensional nature of thermonuclear flashes have been 
two-dimensional \citep{fryxell82a, steinmetz92, shankar92, shankar94,
livne93, glasner95, glasner97, khokhlov95, boisseau96, kercek98}, but
there have been a few three-dimensional studies \citep{khokhlov95,
khokhlov97, garcia99, kercek99}.  Recent progress includes a detailed
two-dimensional study of helium detonations on the surface of a
neutron star that confirmed that a detonation can spread burning over
the entire surface on a time scale consistent with burst rise
times~\citep{zingale01}, three-dimensional simulations of thermonuclear
explosions of Chandrasekhar-mass carbon-oxygen white dwarfs~\citep{reinecke99}, and 
studies of the cellular structure of detonation fronts in both two and three 
dimensions~\citep{timmes00a,timmes01}.

\subsection{The Role of Fluid Instabilities in Flash Problems}

Fluid instabilities and subsequent mixing are expected to play a
fundamental role in the events involving thermonuclear flashes.  For
example, determining whether or not there is substantial mixing
between the accreted hydrogen-helium envelope and the carbon-oxygen 
surface layer of the white dwarf is crucial
to understanding the nova mechanism.  
Mixing of intermediate-mass elements into the accreted layer is
critical because otherwise hydrogen burning would be too slow
to produce a nova.
Furthermore, without this mixing it is difficult to produce the observed
abundances of intermediate-mass nuclei in the ejecta~\citep[see][and references
therein]{rosner01}.  Such mixing may
occur during the first phases of the accretion cycle or when convection in
the accreted layer works its way down to the interface with the white
dwarf. In the latter case, convective undershoot may dredge up
material from the white dwarf and mix it into the accreted 
layer \citep{livio91}. Although some progress has been made
on modeling this nova mixing mechanism \citep{glasner97,kercek98,kercek99}, 
a consensus has not been reached. 
Another proposed mixing mechanism, which is the subject of ongoing
research, is breaking of nonlinear resonant gravity waves 
at the carbon-oxygen surface \citep{rosner00,rosner01,alexakis01,alexakis02}.

In the case of a neutron star, penetration into the crust
by convection in the accreted layer is strongly inhibited by the large
jump in atomic weight between the heavy element crust and the
hydrogen-helium composition of the accreted layer.  Thus, no significant
mixing by this sort of convective dredge-up is expected to occur in
thermonuclear flashes involving the surface layers of a neutron star.
Some mixing may occur via other mechanisms such as shear instabilities, 
but, unlike the nova case, the mixing of the inert heavy elements from the 
underlying neutron star into the accreted layer is not likely to have 
a significant effect on the evolution.
  
Fluid instabilities and mixing are also expected to play a key role
in the explosion mechanism of a Type Ia supernova.  A subsonic burning
front that begins near the center of a massive white dwarf is subject
to Kelvin-Helmholtz, Landau-Darrieus, and Rayleigh-Taylor instabilities
\citep{khokhlov97, khokhlov01, hillebrandt01}.  Growth
of these instabilities dramatically increases the surface area of the
burning front. 
This increase in surface area increases 
both the burning rate and the speed of the front. The dependence of 
the speed of the burning front on fluid instabilities is one of the reasons 
a study of Rayleigh-Taylor instabilities is a key component in our efforts at
V\&V.

Because fluid instabilities play a fundamental role in thermonuclear flash 
events and  may be probed in reasonably good terrestrial 
experiments, experiments involving fluid instabilities have been the focus of our 
validation efforts thus far. In what follows, after describing
the methodology of V\&V and briefly describing 
our numerical methods,  we present verification tests and the results of two validation 
problems-- a laser-driven shock propagating through a multi-layer target and 
the classic Rayleigh-Taylor problem.

\subsection{The Process of Verification and Validation}

The testing step of code development, which involves verification and validation 
of the numerical methods and resulting simulations, is one of 
the most important procedures required for successful numerical modeling. This fact 
has been frequently overlooked in astrophysics. In our description of V\&V, we take 
the point of view that the code to be tested is either one under 
development or an existing code that is being applied to a new problem. The details of 
testing the code will apply in either case. The results of testing will feed back to 
the choices of numerical methods for addressing the physics. If a particular method 
or model fails a test, then another must be chosen. We note that 
verification and validation are necessary but not sufficient tests for determining 
whether a code is working properly or a modeling effort is successful.
These tests can only determine for certain that a code is not working properly.

V\&V is a maturing area of study in the field of CFD, and there is a wealth of 
information available in the 
literature~\citep[cf.][]{aiaa98,oberkampf98,pilch01,roache98a,roache98b}. 
The fundamental strategy of V\&V is the assessment of error and uncertainty in
a computational simulation. The requisite methodology is complex because it must
address sources of error in theory, experiment, and computation. Because these
areas of study present diverse perspectives (and because V\&V is still a developing field),
it is common to find disagreement in the terminology of V\&V~\citep{aiaa98}. 

We adopt the following definitions from the American Institute of Aeronautics 
and Astronautics (AIAA)~\citep{aiaa98}:
\begin{quotation}
\begin{trivlist}
\item Model: A representation of a physical system or process intended to 
enhance our ability to understand, predict, or control its behavior.
\item Modeling: The process of construction or modification of a model.
\item Simulation: The exercise or use of a model. (That is, a model is used
in a simulation)
\item Verification: The process of determining that a model implementation
accurately represents the developer's conceptual description of the
model and the solution of the model.
\item Validation: The process of determining the degree to which a
model is an accurate representation of the real world from the perspective
of the intended uses of the model.
\item Uncertainty: A potential deficiency in any phase or activity of the modeling
process that is due to lack of knowledge.
\item Error: A recognizable deficiency in any phase or activity of modeling that is
not due to lack of knowledge.
\item Prediction: Use of a CFD model to foretell the state of a physical system under
conditions for which the CFD model has not been validated.
\item Calibration: The process of adjusting numerical or physical modeling parameters
in the computational model for the purpose of improving agreement with experimental
data.
\end{trivlist}
\end{quotation}
\citet{roache98a} offers a concise, if informal, summary:
\begin{quotation}
First and foremost, we must repeat the essential distinction between
Code Verification and Validation.
Following \citet{boehm81} and \citet{blottner90}, we adopt the
succinct description of ``Verification" as ``solving the equations right",
and ``Validation" as ``solving the right equations". The code author defines
precisely what partial differential equations are being solved, and convincingly
demonstrates that they are solved correctly, i.e. usually with some order of accuracy, 
and always consistently, so that as some measure of discretization (e.g. the mesh increments)
$\Delta \rightarrow 0$, the code produces a solution to the continuum equations; this
is Verification. Whether or not those equations and that solution bear any relation
to a physical problem of interest to the code user is the subject of Validation. 
\end{quotation}
Roache goes on to point out that in a meaningful sense, 
a ``code" cannot be validated, but only a calculation or range of calculations can 
be validated. He also makes the distinction between verifying a code and
verifying a calculation, noting that ``use of a verified code is not enough."
In our discussion, we will adhere to these definitions as closely as possible.

Another term requiring discussion is ``convergence." In CFD, the  
term is used in two different ways. ``Grid convergence" refers to 
the convergence of the discretization error of the numerical solution
as the grid size and time step approach zero. ``Iterative convergence" 
refers to the convergence of the results of successive
steps of an iterative procedure within a numerical method. 
\citet{roache98a} notes that ``inadequate iterative convergence
will pollute grid convergence results" and that the issue of iterative 
convergence can blur the distinction between verifying a code and verifying a 
calculation because iterative tuning parameters can be problem
dependent. 

\begin{figure}
\epsscale{1.0}
\plotone{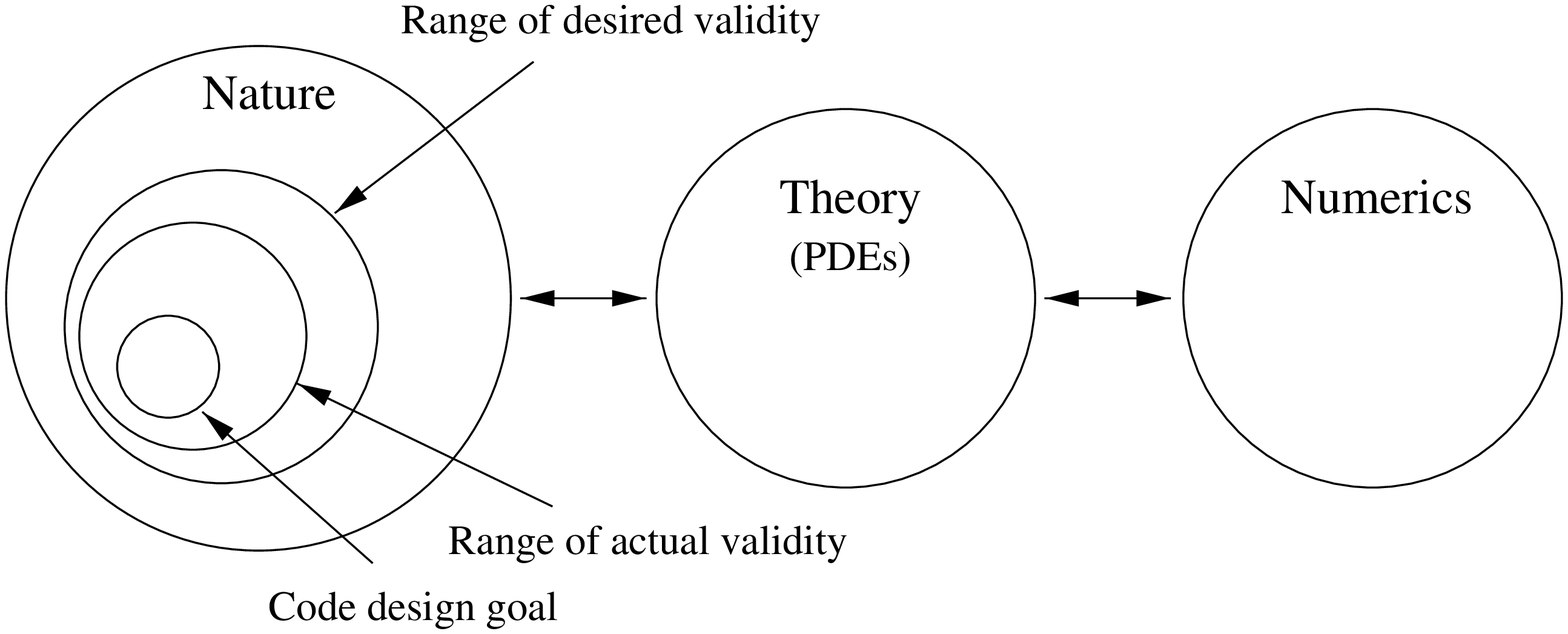}
\epsscale{1.0}
\caption{Schematic of the ranges of validity of a simulation code.
The goal of computational science is to accurately describe Nature 
with a theory implemented by a numerical method. In practice, there is
a desired range of Nature that is to be described, consisting of the 
problems of interest. The  goal of validation is to confirm that the range of actual 
validity of the code and models adequately describes the desired 
range of validity. Verification is testing that a numerical implementation accurately 
represents the model. In this schematic, validation may be thought of as probing the range of actual
validity, represented by the circle labeled range of actual validity. Verification may be
thought of as confirming the mapping between the theory and numerics circles.
\label {fig:valscheme}}
\end{figure}

Figure~\ref{fig:valscheme}, a Venn diagram~\citep{venn80}
illustrating the parts of numerical modeling, provides a schematic for considering 
the role of verification and validation in numerical modeling. On the left, the largest 
circle represents Nature, or at least the part of Nature in which the problem
of interest resides. The smaller circles inside of Nature represent, from
largest to smallest, the range of desired validity of the code, the 
range of actual validity, and the range of the design goals of the code.
The circle in the center of the diagram (to the right of the Nature circle)
represents the theory describing Nature (which may be thought of 
as the model), and the circle to the right represents 
numerical scheme(s) implementing the theory (modeling and simulation).

Verification begins by identifying the purported design goal of the code 
or code modules, which is represented by the smallest circle in the center of 
the Nature circle on the schematic. This is the step Roache refers
to as defining ``precisely what partial differential equations are being solved,"
though the process may require a larger scope. In the language of the AIAA, this may
be thought of as identifying and understanding the implementation of the model.
Verification, confirming that
simulations produced by the code accurately implement the design goal of the 
code (i.e. the implementation accurately represents the model 
and the solution to the model),
may be thought of as confirming the ``mapping" between the theory and numerical 
method circles on the schematic. 
The process requires identification and a quantitative 
description of the error, and the strategy is typically a systematic study of mesh 
and time step refinement as is appropriate for finite difference, 
finite volume, and finite element methods. Other numerical methods such as vortex, 
lattice gas, and Monte Carlo methods require different procedures~\citep{aiaa98}.

Verification tests are typically simple problems, often with known 
analytic solutions, that can be used to study the accuracy and
convergence rates of a code.
Despite the relative simplicity of the tests, verification is a difficult
process. Analytic solutions, which are easiest to compare against, 
are few and usually are limited tests of the physics.
Even in these simple cases, singularities and discontinuities complicate
the process of verification. Singularities arising from the geometry or
the coordinate system should be removed where possible.
Singularities inherent in the conceptual model (e.g. shocks and discontinuities
in the flow) require special attention.
These structures will typically exhibit a lower order of 
convergence from the rest of the flow, if indeed they lead to a converged solution
at all.
An added level of complexity arises in the case of discontinuous flows
because not all numerical methods 
for solving a given set of PDEs are equivalent. Different methods handle 
flow discontinuities
differently and/or rely on different (intrinsic numerical or physically-motivated)
subgrid models. Therefore, confirming that a numerical method is correctly solving 
a set of PDEs can also be a validation problem.

Test problems, even without analytic solutions, may be devised to 
monitor conservation, symmetry properties, and effects of boundary conditions~\citep{aiaa98}. 
Additional progress may be made by testing self-convergence.
Self-convergence can be difficult to test, however, and 
changing the resolution by a only factor of two does not demonstrate a 
converged answer~\citep{fryxell94}.
Further, resolving the length and time scales relevant to 
the physical problem well enough for convergence may be 
prohibitively expensive.
Also, simulations with complex physics, particularly three-dimensional simulations,
may not have resolved solutions~\citep{aiaa98}.
We further note that convergence tests say nothing about the correctness of
the answer in that it is possible for solutions to converge to the wrong answer. 

Verification testing also may include code-to-code
comparisons, that is, the comparison of the results of simulations
performed with different codes. For these types of comparisons, 
care should be taken to find accurate, benchmarked solutions calculated very 
carefully by independent investigators,
preferably using different numerical approaches~\citep{aiaa98}.
Examples in astrophysics and cosmology of code-to-code comparisons
include the Geospace Environmental 
Modeling (GEM) Magnetic Reconnection Challenge~\citep{birn01} and
the Santa Barbara Cluster Comparison Project~\citep{frenk99}. 
The GEM Reconnection Challenge is a collaborative study of phenomenon
of magnetic reconnection, which plays an important role in magnetosphere dynamics.
The Santa Barbara Cluster Comparison Project compared the results of
twelve codes simulating the formation of a galaxy cluster in a flat
cold dark matter universe.  Both studies performed simulations starting
from a uniform set of initial conditions, with the goal of studying how
different numerical methods capture the behavior of the system under
study.  Diagnostics were constructed that were sensitive to the physical 
processes essential to these systems and robust enough to
compare across codes. 

The utility of code-to-code comparisons both for finding bugs and for
increasing the understanding of the behavior of different numerical
methods in complex situations is clear. This procedure can only be 
meaningful, however, if the codes involved have undergone other rigorous
V\&V tests.  As with code-to-analytic or code-to-experimental
comparisons, these tests can only indicate possible problems.
Code-to-code comparisons can also include consistency tests such as the
nightly comparison tests that FLASH undergoes, which will be discussed
below.

Validating a simulation requires identifying the key elements involved in
the simulations and for each element (as well as the integrated code)
constructing test problems that have the results of laboratory 
experiments as the accepted results. 
Key elements include parts of the code that describe the fundamental physical 
processes and include items such as transitions to low and high Mach numbers,
transport of energy by conduction or radiation, transport of energy by
advection (convection), source terms (e.g. nuclear burning), equations of state,
opacities, and microscopic transport such as molecular diffusion and viscosity.
Validation problems tend to be much more complex than verification problems
and typically have no analytic solutions. These problems nevertheless involve phenomena 
that are sufficiently simple to be studied both by experiment and 
simulation, and these problems are used to determine 
whether a particular calculation reproduces the outcome of the 
phenomenon or experiment. 

We note that validation goes beyond purely numerical testing; it 
includes testing the fundamental assumptions and concepts that go into a model
and probing the range of validity of a model.
In the schematic, validation may be thought of as probing the area of
the actual validity circle. An example that has implications for the problems here
is that of compressibility.  As \citet{roache98a} mentions, a thoroughly 
verified incompressible fluid dynamics code will produce invalid results when 
applied to a problem in which compressibility of the fluid affects the dynamics. 
The situation is more complicated, however, if one considers the validity 
of applying a compressible code to an incompressible problem (or a very low Mach 
number flow). In this case, the compressible code may correctly address the 
problem, but the CFL limit will require a huge number of very small time steps, making 
the problem intractable and the solution less accurate due to the 
accumulation of numerical errors.
Thus incompressible (or very low Mach number) flow may be within the formal 
range of validity of a compressible code (i.e., the range that is 
mathematically well-posed), but not within the actual or practical range of validity.

The challenges associated with validating an astrophysical simulation code 
exceed those of validating a standard fluid dynamics code. 
Astrophysical events are often complex and involve many interacting physical
processes, each of which must be tested. 
Validating any simulation code with laboratory experiments can be a difficult
process, particularly if the diagnostic resolution of the experiment is 
poor, if there are significant uncertainties in the material properties, or 
if the initial/boundary conditions are not well-defined.  
The situation is especially acute in astrophysics, 
where we are limited to observations of distant objects.
The interiors of stars, for instance, do not 
lend themselves to direct observation, and, even if this were not
a problem, the vast majority of astrophysical objects are too
far away to resolve.
Observations of thermonuclear flashes can only show the results of the 
events (light curves and spectra) and not the details of initiation of the outburst.
Also, the length scales of the astrophysical 
objects present challenges. Flows within stars, for example, are expected
to occur at Reynolds numbers 
greater than 10$^9$, and 
terrestrial experiments cannot approach such a regime.

In addition, validation testing includes systematic grid sensitivity studies to 
asses grid convergence error and assess the level of refinement 
necessary to capture the key physical effects~\citep{aiaa98}.
Accordingly, many of the difficulties in astrophysical verification
also appear in validation. As noted, the flows of interest in 
astrophysics, particularly flows in thermonuclear flash events, 
involve unsteady flows, and convergence testing is more difficult for
these flows. The required accuracy of validation activities, however,
is not generally as stringent as that of verification activities~\citep{aiaa98}. 

Another issue making astrophysical validation more complicated than
CFD validation is the choice of equation of state. The CFD literature we
studied mentions equations of state as possible sources of error, but
does not emphasize the importance of choice of equation of state on 
the validity of the model. Phenomena such as degeneracy and other
interactions play a critical
role in astrophysics, and the influence of these enter the simulation
through the equation of state. 
The properties of an equation of state (e.g. convexity) are particularly
important when using hydrodynamics methods that solve the Riemann problem
as does the principal hydrodynamics module in FLASH~\citep{menikoff89}.
Considerable effort can go into validating a
hydrodynamics code with a known, accurate equation of state, 
but coupling it to an inappropriate or inaccurate equation of
state will invalidate any simulations. Also, effort expended in testing
equations of state relevant for laboratory experiments will 
not necessarily improve confidence in predictions made by astrophysical
simulations that require another equation of state.

Because of these complexities, it seems unlikely that 
one could ever satisfactorily validate an astrophysical simulation. Instead,
validation efforts focus on laboratory experiments that capture 
the relevant physics, with the expectation that the experience gained
from these closely related cases builds confidence in the predictions
of the astrophysical simulations.  Accordingly, a significant part of the 
challenge of validating astrophysical simulation codes is to find acceptable 
laboratory experiments.  A good experiment for validation should be a good 
experiment itself (that is, it should provide accepted, repeatable results), 
it should adequately capture a significant portion of the physical processes 
of interest, and it should be diagnosed well enough for a meaningful comparison 
to simulation. 

A final subject to describe in the methodology of V\&V is calibration. Calibration
is not validation. Instead, calibration is a process performed in order to improve the
agreement of computational results with experiments, and calibration does not
generate the same level of predictive confidence as validation.
Calibration is performed when there is uncertainty in the modeling of complex
processes and also when there are incomplete or imprecise measurements in the
experiments. Calibration involves adjustments to parameters
in subgrid models, reaction rates, and boundary conditions, and it includes
assumptions about minimal or optimal levels of mesh refinement that
are made in cases where there are not completely resolved solutions~\citep{aiaa98}. 

We complete our introduction with the observation that 
much of what we have said about simulation validation is true
of validating any theoretical model of a system, including
more traditional analytic models. 

\section{Numerical Method}

The FLASH code is a parallel, adaptive-mesh
simulation code for studying multi-dimensional compressible reactive 
flows in astrophysical environments. 
It uses a customized version of the
PARAMESH library \citep{macneice99,macneice00} to manage a 
block-structured adaptive grid, adding resolution elements in areas
of complex flow.  
The current models used for simulations
assume that the flow is described by the Euler equations for compressible, inviscid
flow. FLASH regularizes and solves these equations by an explicit, 
directionally split method (described below), carrying a separate advection equation 
for the partial density of each chemical or nuclear species as required for reactive 
flows. The code does not explicitly track interfaces between fluids so some
numerical mixing can be expected during the course of a calculation.
FLASH is implemented mostly in Fortran 90 and
uses the Message-Passing Interface library \citep{gropp99} to achieve portability.
FLASH makes use of modern object-oriented software technology
that allows for minimal effort to swap or add physics modules.
Accordingly, the development of FLASH requires development and
testing of each module as well as development and testing
of the framework integrating the modules. In the subsections below,
we provide details of some of the modules in FLASH.
Complete details concerning the algorithms used in the code, the structure of 
the code, selected verification tests, and performance may be found
in \citet{fryxell00} and \citet{calder00}. 

\subsection{Hydrodynamic Module}

The primary hydrodynamic module in 
FLASH is based on the PROMETHEUS code \citep{fryxell89}
and evolves systems described by the Euler equations 
for compressible gas dynamics in one, two, or three dimensions. 
The evolution equations are solved using a modified version 
of the Piecewise-Parabolic Method (PPM), which is described in detail in
\citet{woodward84} and \citet{colella84}.
PPM is a shock capturing scheme in which dissipation is
used to regularize the Euler equations.
(See \citet{majda84} for a discussion
of the importance of dissipative mechanisms.)
Modifications to the method include the capability to use general equations of 
state \citep{colella85}. 
PPM is a higher-order version of the method developed by~\citet{godunov59,godunov61}.
Godunov methods are finite-volume conservation schemes that 
solve the Riemann problem at the interfaces of the control
volumes to compute fluxes into each volume. The conserved fluid 
quantities are treated as cell averages that are updated by the fluxes 
at the interfaces. This treatment has the effect of introducing explicit 
non-linearity into the difference equations and permits the calculation of 
sharp shock fronts and contact discontinuities without introducing 
significant non-physical oscillations into the flow.  
The original Godunov method is limited to first-order accuracy in both space 
and time because the distribution of each variable in each
control volume is assumed to be constant. 
PPM extends this method by representing the flow variables as 
piecewise-parabolic functions and also by incorporating monotonicity
constraints to limit unphysical oscillations in the flow.
PPM is formally accurate to only second order in both space and 
time, but performs the most critical steps to third- or fourth-order 
accuracy. This results in a method which is considerably more accurate and
efficient than most second-order codes using typical grid sizes.
A fully third-order (in space) method provides only a slight
additional improvement in accuracy but results in a significant
increase in the computational cost of the method. 
 
PPM is particularly well-suited to flows involving discontinuities
such as shocks and contact discontinuities. The method also performs
well for smooth flows, although other schemes that do not
perform the additional steps for the treatment of discontinuities
are more efficient in these cases.  The high resolution and
accuracy of PPM are obtained by the explicit non-linearity of the
scheme and through the use of smart dissipation algorithms, which are
considerably more effective at stabilizing shock waves than the more
traditional explicit artificial viscosity approach. Typically,
shocks are spread over only one to two grid points, and post-shock 
oscillations are virtually nonexistent in most cases.  Contact
discontinuities and interfaces between different fluids create special
problems for Eulerian hydrodynamics codes.  Unlike shocks, which
contain a self-steepening mechanism, contact discontinuities spread
diffusively during a calculation; they continue 
to broaden as the calculation progresses.  PPM contains an
algorithm that prevents contact discontinuities from
spreading more than one to two grid points, no matter how far they
propagate.

The PPM implementation in FLASH is a directionally split,
Direct Eulerian formulation. The hydrodynamics module applies
the PPM method in one-dimensional sweeps across a block of data,
advancing the time two steps. Reversing the order of the sweep
for the second time step preserves second-order accuracy in time~\citep{strang68}.
In three dimensions, the sweeps are performed in the order $xyz-zyx$
(for Cartesian geometry). The algorithm uses a nine-point stencil
in each direction, requiring that each block have four ghost zones
on each side. In addition, a small multi-dimensional 
artificial viscosity is added to provide a weak coupling between
adjacent rows and columns in the directionally split scheme.

\subsection{Source Terms}

FLASH incorporates source terms that are operator split with the
hydrodynamics evolution. Two of these are modules for self-gravity and
thermonuclear burning. The gravitational module computes the source term
(acceleration and/or potential) for the effects of the force of gravity, 
which in the case of an astrophysical
object such as a star cannot be treated as an applied external field.
The main role of the thermonuclear burning module in FLASH 
is to provide the magnitude and sign of the energy generation rate.  
A secondary role is to evolve the abundances of the nuclear species.

The gravitational module solves the Poisson equation for the
gravitational potential. We have implemented both 
multigrid \citep{martin96} and multipole methods for the 
solution on our adaptive mesh.
We have incorporated methods for periodic and isolated boundary conditions.

Thermonuclear energy generation is typically the largest source or
sink of energy in regions conducive to nuclear reactions; so accurate
determination of the energy generation rate is essential to obtaining
accurate simulations.  Calculating an accurate energy generation rate,
however, is very expensive in terms of computer memory and CPU time.
Decreasing the expense to compute a model
requires making a choice between having fewer isotopes in the reaction
network or having less spatial resolution.  The general response to
this tradeoff has been to evolve a limited number of isotopes and
thus calculate an approximate thermonuclear energy generation rate.
For example, when studying explosive burning in pure helium environments,
a network composed of $^4$He,
$^{12}$C, $^{16}$O, $^{20}$Ne, $^{24}$Mg, $^{28}$Si, $^{32}$S,
$^{36}$Ar, $^{40}$Ca, $^{44}$Ti, $^{48}$Cr, $^{52}$Fe, and $^{56}$Ni
is usually sufficient. This minimal set of nuclei, usually called an 
$\alpha$-chain network, can return an energy generation rate that is generally 
within $\sim$ 30\% of the energy generation rate given by much larger nuclear
reaction networks \citep{timmes00b}.

Even with a reduced set of nuclei in the reaction network, it is
desirable to solve the reaction network equations as efficiently as
possible because there can be over 10$^9$--10$^{12}$ calls to the thermonuclear
burning modules in typical two- and there-dimensional hydrodynamic simulations of 
astrophysical flashes.  \citet{timmes99} compared a variety of methods for
solving the stiff system of ordinary differential equations that
constitute a nuclear reaction network. The results of this
study led to the choice of methods included with the standard FLASH 
distribution \citep{fryxell00}.  

\subsection{Equations of State}

FLASH includes two equations of state in its standard distribution, a
gamma-law equation of state and a tabular Helmholtz free energy equation 
of state for stellar interiors. The gamma-law equation of state models a simple
ideal gas with a constant adiabatic index. Simulations are not
restricted to a single ideal gas, however, because the code allows for 
simulations with several species of ideal gases with different gammas. 
While this equation of state executes very efficiently because of 
its simplicity, it is limited in its range of applicability for
astrophysical flash problems. The stellar equation of state includes
contributions from blackbody photons, completely ionized nuclei, and 
degenerate/relativistic electrons and positrons.  
A thermodynamically consistent interpolation of the
Helmholtz free energy (which satisfies the Maxwell relations exactly)
is used for the electron-positron contribution.
This stellar equation of state has been subjected to considerable
analysis and testing \citep{timmes00c},
and particular care was taken to reduce the numerical error introduced by the
thermodynamical models below the formal accuracy of the hydrodynamics
algorithm \citep{fryxell00,timmes00c}.
In addition to these, we are testing additional equations of state and describe 
below the process of verifying these in the context of the hydrodynamics 
algorithm.

\section{Verification Tests}

We have an entire suite of test problems that are 
regularly applied to the code for software verification.
These are standard test problems in the field of 
fluid dynamics, and many of these problems have analytic solutions. 
The remaining problems produce well-defined flow features that make for 
stringent tests of the code. These test problems are run on a wide 
variety of platforms and compilers at weekly or more frequent intervals, 
and the results are compared to accepted earlier results. For a code with 
many developers, such stringent testing is crucial for having faith in 
the results obtained by the code. The results of the tests are posted, and  
any deviation of the results from the ``good" previously computed
results is noted so that cause of the  deviation may be investigated.
These code-to-code comparisons (here the two codes are different versions
of the same code) allow any errors introduced to be spotted immediately
and have been invaluable for the development of FLASH.

As of this writing, the test suite includes

\begin{itemize}

\item The strong shock tube problem of \citet{zalesak00}.
This test is more stringent than the usual Sod test (described below)
because of the stronger discontinuities across the shock interface and 
the narrow density peak that forms behind the shock.

\item The Sedov explosion problem \citep{sedov59}, a purely hydrodynamical 
test involving strong shocks and non-planar symmetry. The problem 
consists of the self-similar evolution of a cylindrical or spherical blast 
wave from a delta-function initial pressure perturbation in an otherwise
homogeneous medium. In practice, the explosion is initiated by depositing a 
quantity of energy into a small region at the center of the 
computational grid. The profile and speed of the resulting expanding blast 
wave are verified by comparison to the analytic solution. 

\item The interacting blast wave problem. Originally used by 
\citet{woodward84}, this problem tests the ability of a
hydrodynamics method to handle strong shocks. 
It has no analytic solution, but since it is one-dimensional it is 
easy to produce a numerically converged solution by running the code with a 
very large number of zones, permitting an estimate of the self-convergence 
rate when discontinuities are present. For FLASH it also
provides a good test of the adaptive mesh refinement scheme.

\item A wind tunnel with a step \citep{emery68}. Although it also has no 
analytic solution, this problem exercises the ability 
of a code to handle unsteady shock interactions in multiple dimensions. 
It also serves as a test problem with irregular boundaries.

\item A shock forced through a jump in mesh refinement. The mesh refinement 
algorithm in FLASH is designed to avoid this situation, but there may be 
cases when it is desirable to force a jump in the mesh refinement. 
Such cases may arise from limited computational resources, or from 
carrying regions where a fully refined solution is not necessary. The
test monitors the ability of the code to handle such a situation. 

\end{itemize}
Additional tests, some specific to certain
problems, are also regularly administered to the code.
Details and results of some of these tests may be found in \citet{fryxell00}.
A gallery of results of verification tests 
may be found at \url{http://flash.uchicago.edu/}, as well as updates
to the current suite of test problems. 

As FLASH develops and new physics modules are added, we expand the
test suite to include tests of the new modules that typically
involve source terms. Verification testing of the nuclear burning 
module in FLASH has consisted of testing the network in use against 
larger networks and testing flame speeds against speeds obtained from
other methods. Results of FLASH simulations indicate that we match
the flame speeds found by \citet{timmes92}. Complete details of the 
flame speed verification tests will be reported with the results of 
a study of flames and flame-vortex interactions \citep{zingale02b}.
Verification tests of the gravitational modules in FLASH include
a homologous dust collapse \citep{colgate66,monch89}, a collapsing 
isothermal gas sphere \citep[and references therein]{lai00},  
a two-dimensional problem consisting of Gaussian density peaks at
different locations and with different widths \citep{huang00},
and the Jeans instability \citep{jeans02,chandra81}.
Details and results of these gravitational tests will appear in
a forthcoming report \citep{ricker01}.

\subsection{Hydrodynamics Module}

In this section we present more extensive verification testing of the 
hydrodynamics algorithm. The selected problems include tests of pure 
advection, sound wave propagation,
and shocks. Advection problems in one dimension test the ability of the
code to maintain the shape of a density pulse propagating at a constant
velocity across the mesh, thereby testing the treatment of flow features
that move at characteristic speeds of the hydrodynamics equations.
Noise generated by a feature will move with the feature, accumulating
as the calculation advances, making these sensitive tests. 
Advection problems similar to these were first proposed
by \citet{boris73} and \citet{forester77}. 
We also consider the advection of an isentropic vortex \citep{shu98,yee00}.
This two-dimensional problem exposes the directional splitting
of the hydrodynamics algorithm to scrutiny.
The sound wave test consists
of testing the ability of the code to maintain the shape of
a simple sinusoidal sound wave propagating across the mesh. The problem
is similar to the dispersive sound wave problem of \citet{masset00}.
We test the handling of shocks with the shock tube problem 
of \citet{sod78}, a simple test of the ability of a compressible
code to capture shocks and contact discontinuities
and to produce the correct profile in a rarefaction. This problem
also tests the ability of the code to satisfy correctly the
Rankine-Hugoniot shock jump conditions. We also test the ability
of the code to maintain a stationary shock, which further tests
the Riemann solver.

The principal improvement in these tests over the previously
published results is in the application of the initial conditions. 
The previous work constructed initial conditions  
at the cell centers, i.e. point values. For the new tests, 
thermodynamic quantities were 
interpolated via a higher order method to provide the cell averaged
quantities~\citep[see][]{zingale02a}. This interpolation step provides
initial conditions more consistent with the assumptions of the solution
technique and produces better results. Most of the tests presented here
were performed on a uniform mesh. Verification 
of applications using adaptive mesh refinement, as implemented in FLASH,
is the subject of ongoing research and the literature of this 
problem is sparse. The tests had constant time steps at each 
resolution such that (i) each simulation ended at exactly the same 
evolution time, and (ii) the ratio of $\Delta t / \Delta x$
was constant across different resolutions, corresponding to a
fixed Courant number.

\begin{figure}
\epsscale{0.9}
\plotone{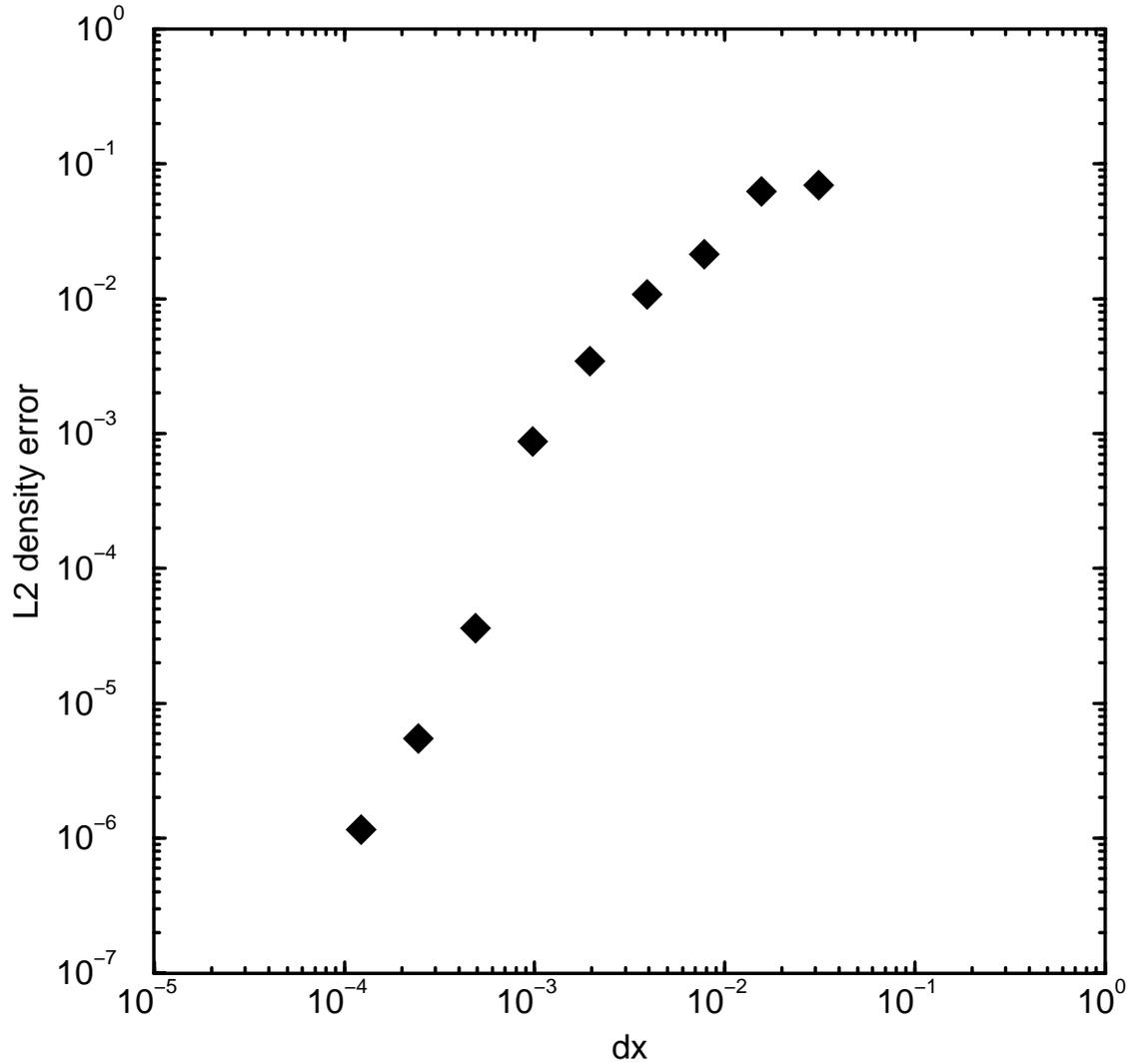}
\epsscale{1.0}
\caption{Plot of L2 norm of the density error vs.\ mesh spacing for a 
simple advection test consisting of a Gaussian density pulse propagating at a 
constant velocity across the mesh. The simulation used full PPM including 
contact steepening.
The sudden decrease in error for resolutions below $\approx 10^{-3}$
is due to 
the narrow pulse being sufficiently well-resolved that contact
steepening was no longer applied. The different behavior of the density
error at the lowest resolutions occurs because the initial conditions are not
adequately resolved.
\label {fig:gaus_cs}
}
\end{figure}

\begin{figure}
\epsscale{0.9}
\plotone{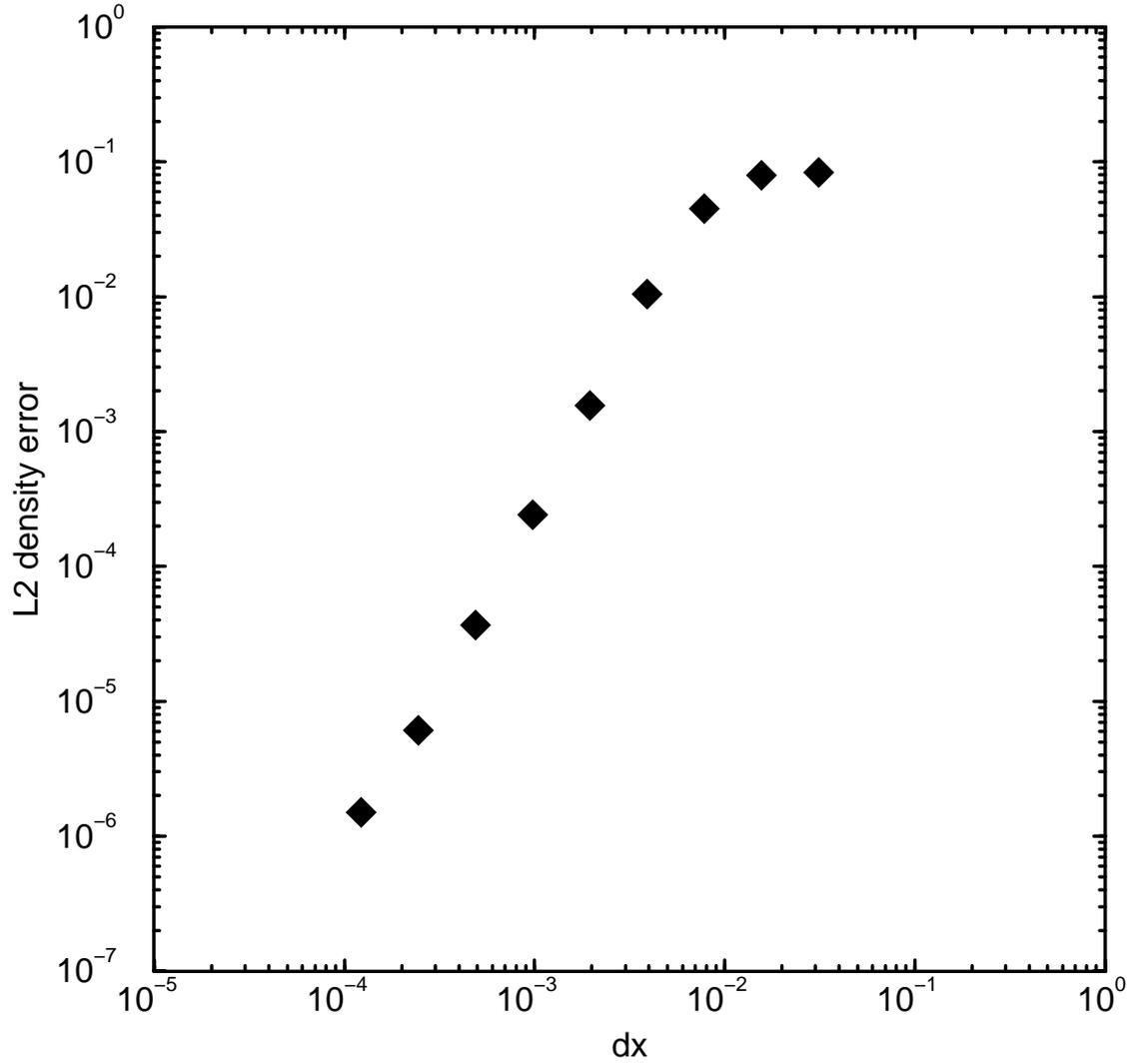}
\epsscale{1.0}
\caption{Plot of L2 norm of the density error vs.\  mesh spacing for a 
simple one-dimensional advection test consisting of a Gaussian density pulse 
propagating at a constant velocity across the mesh. 
Contact steepening was turned off to make sure we understood the
feature of the curve in Figure~\ref{fig:gaus_cs}. The different 
behavior of the density
error at the lowest resolutions occurs because the initial conditions are not
adequately resolved.
\label {fig:gaus_ncs}
}
\end{figure}

The first test consists of a Gaussian pulse propagating across the 
simulation mesh with a constant velocity. 
This problem tests the treatment of narrow
flow features, which may be clipped by the introduction of artificial
dissipation \citep{zalesak87}. 
The initial conditions are a planar density pulse in a region of
uniform, dimensionless pressure $P_0$ and dimensionless velocity $v_0$.
The density pulse is defined via
\begin{equation}
\label {adv1}
\rho(s) = \rho_1\phi(s/w) + \rho_0\left[1-\phi(s/w)\right]\ ,
\end{equation}
where $s$ is the distance of a point from the pulse mid-plane, and $w$ is
the characteristic width of the pulse. The pulse shape function
$\phi$ for a Gaussian pulse is
\begin{equation}
\label {adv3}
\phi_{\rm GP}(\xi) = e^{-\xi^2}\ .
\end{equation}
Two sets of simulations were performed from these initial conditions, one
set with contact steepening and one set without. The simulations ran 
for 0.2 time units
with a fixed Courant number of 0.1. 
Figure~\ref{fig:gaus_cs} shows density error in the L2 norm 
for nine simulations of increasing resolution with contact steepening, 
and Figure~\ref{fig:gaus_ncs}
shows density error in the L2 norm for the equivalent simulations without
contact steepening.
The use of cell-averaged initial conditions led to better results than
those of the previous Gaussian advection test \citep{fryxell00}. 

\begin{figure}
\epsscale{0.9}
\plotone{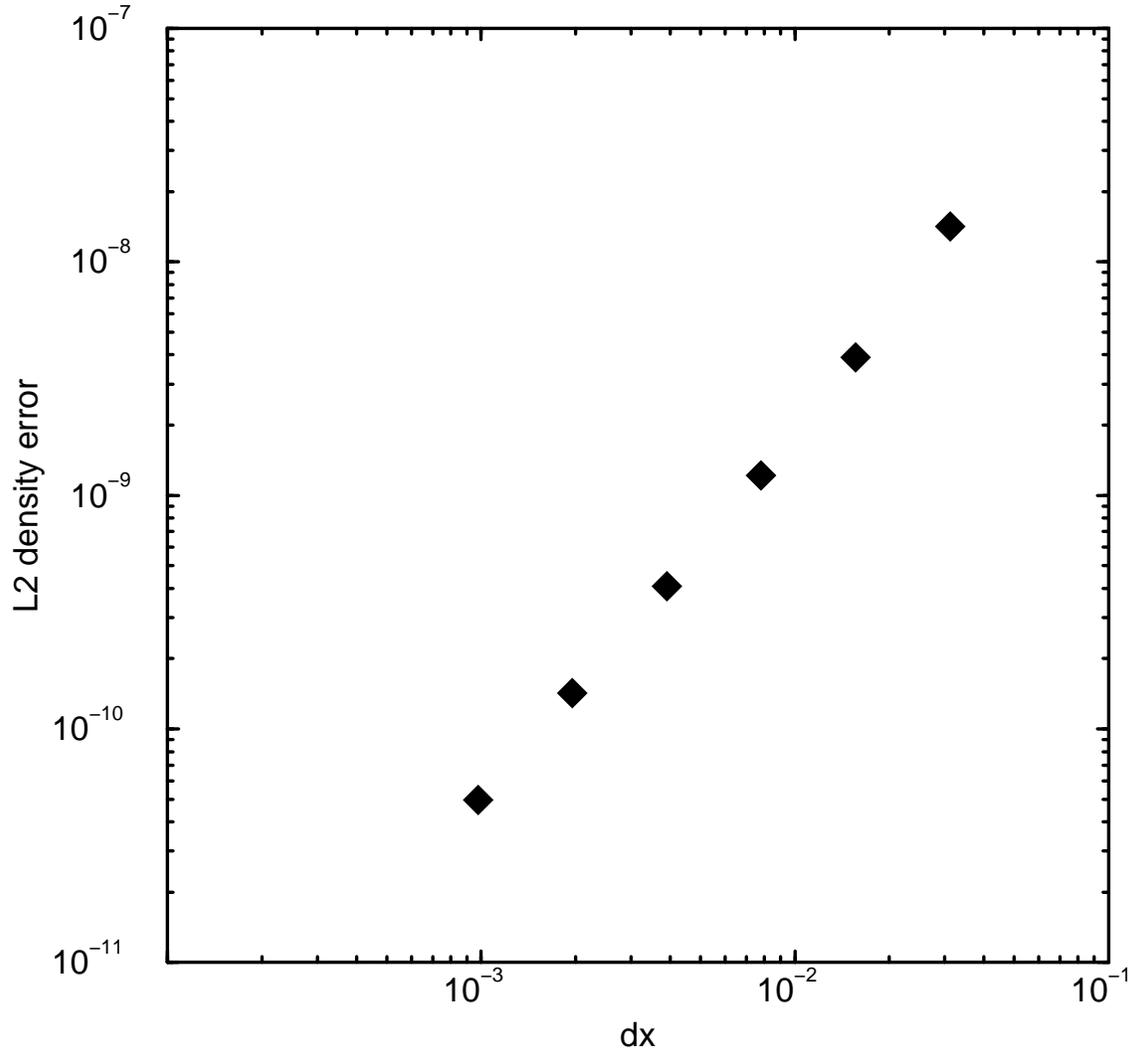}
\epsscale{1.0}
\caption{Plot of L2 norm of the density error vs.\ mesh spacing for one-dimensional simulations
of a sinusoidal sound wave propagating across the simulation domain.
\label {fig:swave}
}
\end{figure}

The second test is the one-dimensional propagation of a sinusoidal sound wave
consisting of a density and pressure perturbation propagating
at the sound speed, $c_s$~\citep{masset00}. The 
background pressure $p_0$ and density $\rho_0$
in arbitrary units are 3.0 and 50.0, and the perturbed
density, pressure, and corresponding velocity are given by
\begin{equation}
\label {sound1}
\rho = \rho_0 + \epsilon \rho_0 \cos(kx), 
\end{equation}
\begin{equation}
\label {sound2}
p = p_0 + c_s^2(\rho  - \rho_0),
\end{equation}
and
\begin{equation}
\label {sound3}
v = c_s\frac{\rho-\rho_0}{\rho_0} ,
\end{equation}
where $\epsilon$ is the amplitude ($10^{-6}$) and
$k$ is the wave number. The formulation corresponds to a rightward
propagating sound wave. 
The sound wave test problem is similar to the Gaussian propagation
problem except that it tests the entire hydro module instead
of just the advection terms.
It has a smooth solution, until the
wave steepens into a shock wave, so it should show the correct
order of convergence for the entire hydro module.
Figure~\ref{fig:swave} shows the L2 norm of the density 
error after the sound wave propagated one wavelength. 

The third test is a stationary shock. This test demonstrates
that the implementation of the Riemann solver is correct,
that is, the Rankine-Hugoniot equations are being correctly
solved and there is no intrinsic numerical viscosity in
the Riemann solver. The initial conditions 
consisted of a configuration similar to the Sod problem (described below), but
in a moving fluid such that the shock should remain stationary
relative to the mesh. The initial pressure discontinuity is
larger than that of the Sod problem, with a $10^5$ pressure jump
across the discontinuity. The result after
4 CFL times agrees exactly with the 
initial conditions.  The presence of intermediate values of the thermodynamic
quantities between the two sides
of the initial conditions after a few time steps would indicate
a failure.

\begin{figure}
\epsscale{0.9}
\plotone{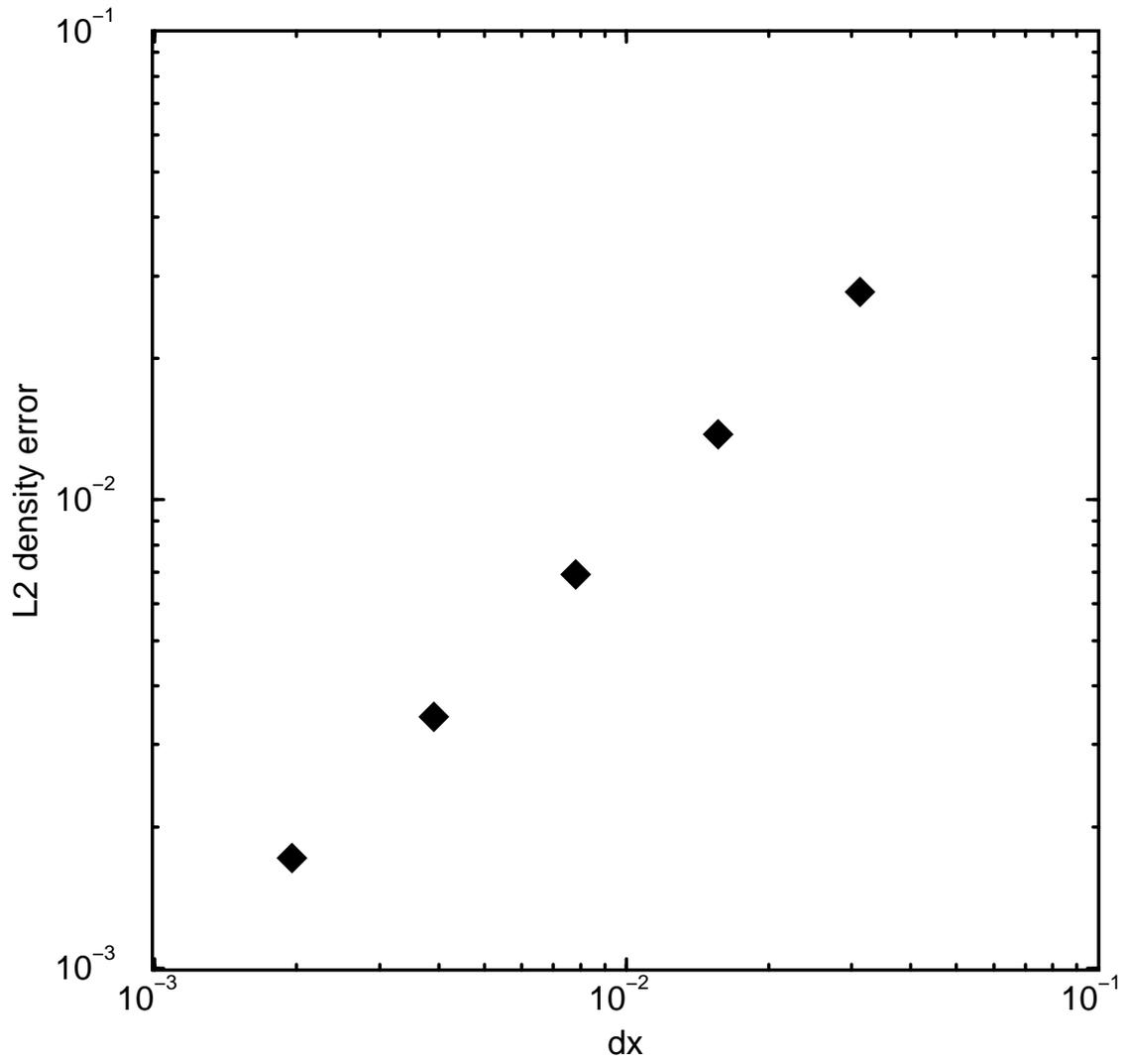}
\epsscale{1.0}
\caption{Plot of L2 norm of the density error vs.\ mesh spacing for the
Sod shock tube test.
\label {fig:sod}
}
\end{figure}

The fourth test is the Sod shock tube~\citep{sod78}, consisting of
a planar interface between two fluid states initially at rest. 
The density and pressure discontinuities across the interface 
produce a flow that develops a shock, a contact discontinuity,
and a rarefaction wave. The initial conditions were
\begin{eqnarray}
\rho_{\mbox{\scriptsize{left}}} & = & 1.0 \\
\rho_{\mbox{\scriptsize{right}}} & = & 0.125 \\
P_{\mbox{\scriptsize{left}}} & = & 1.0 \\
P_{\mbox{\scriptsize{right}}} & = & 0.1
\end{eqnarray}
and the ratio of specific heats, $\gamma, = 1.4\:$.
The simulations for this test were two-dimensional, with the 
flow propagating along the $x$-axis.
Figure~\ref{fig:sod} shows the L2 norm of the density error
for five simulations of increasing resolution at $t = 0.2$.
The curve demonstrates the expected first-order convergence.

\begin{figure}
\epsscale{0.9}
\plotone{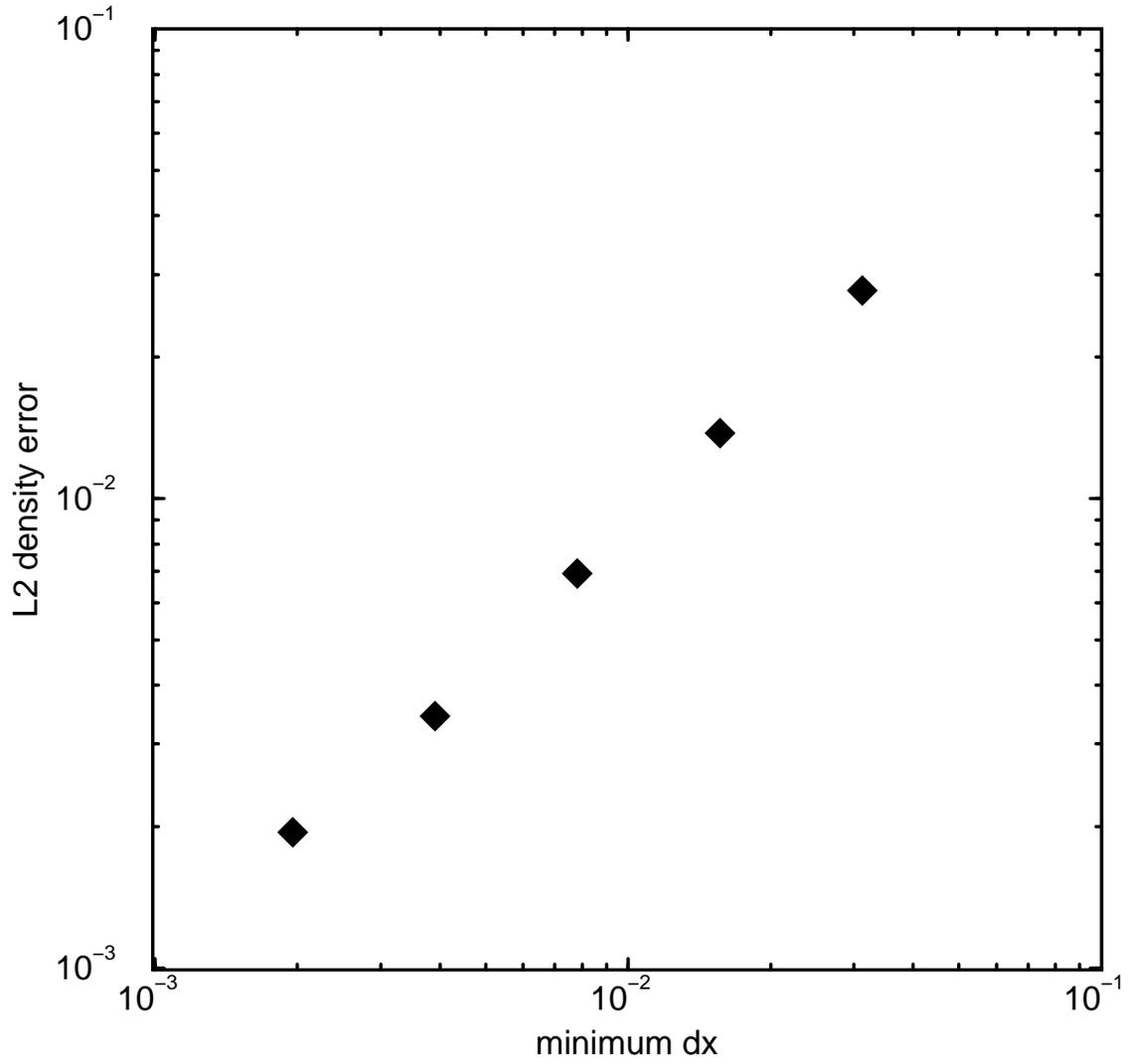}
\epsscale{1.0}
\caption{Plot of L2 norm of the density error vs.\ minimum mesh spacing for the
Sod shock tube test performed on an adaptive mesh. The maximum mesh spacing of
each simulation was that of the least-resolved uniform mesh simulation.
\label {fig:sod_amr}
}
\end{figure}

Because the validation tests presented below 
involve shocks and material interfaces and were performed
on an adaptive mesh, it is appropriate to
provide a minimal, related verification test for an adaptive mesh.
To our knowledge, the literature on verifying a 
block-structured adaptive mesh application is largely non-existent, 
and even such terms as mesh convergence are poorly defined. 
We will address this issue in future work.
For this work, we performed a study analogous to the previous 
Sod test on an adaptive mesh. The Sod problem, with its 
shock and contact discontinuity, is an appropriate 
test of our adaptive scheme, which refines or de-refines the 
mesh in regions in which the second derivatives of hydrodynamic 
variables (by default, density and pressure) 
is larger or smaller than some threshold~\citep{fryxell00}. 
The effect is to add or remove resolution in the simulation.
These criteria are the same as those used in the validation tests.
Figure~\ref{fig:sod_amr} shows the L2 norm of the
density error for five adaptive mesh simulations corresponding
the the uniform mesh simulations above. In this case, the $x$-axis
is the finest resolution of the adaptive mesh. The coarsest resolution
of each simulation was that of the lowest resolution uniform mesh
simulation. The results show the expected first-order convergence.

The final test simulates the advection of an isentropic vortex in
two dimensions~\citep{shu98,yee00}. In our test, the vortex
propagates diagonally with respect to the grid.
This problem is chosen because its solution is smooth
and an exact solution is available.
The simulation domain is a square, $-5.0 \leq x, y \leq 5.0$.
The ambient conditions are, in non-dimensional units,
$u_{\infty}=1.0$, $v_{\infty} = 1.0$, and $T_{\infty} = 1.0$.
The following perturbations are added to get $u$, $v$, and $T$ fields:
\begin{eqnarray}
\delta u &=&
-y \frac{\beta}{2 \pi} \exp \left( \frac{1-r^2}{2} \right) \\
\delta v &=&
x \frac{\beta}{2 \pi} \exp \left( \frac{1-r^2}{2} \right) \\
\delta T &=&
- \frac{(\gamma - 1 ) \beta}{8 \gamma \pi^2} \exp \left( {1-r^2}
\right) 
\end{eqnarray}
where $r^2 = x^2 + y^2$, $\gamma=1.4$ is the ratio of
specific heats and $\beta=5.0$ is a measure of the vortex strength.
The density is then computed by $\rho = (T_\infty + \delta
T)^{1/(\gamma-1)}$.
The conserved variables (density, $x$- and $y$-momentum, and total energy)
can be computed from the above quantities.
The flow field is initialized by computing cell averages of the
conserved variables: each average is approximated by averaging
over $10^2$ subintervals in the cell.

For all isentropic vortex simulations, periodic boundary conditions 
were applied, the
time step was fixed and the Courant number was approximately~0.94, and
the error was calculated at time $t=2.0$. Solutions were obtained on
$40^2$, $80^2$, $160^2$, $320^2$, $640^2$, and $1280^2$ equispaced
meshes. The exact solution at time $t$ is the initial condition
translated by $(u_{\infty} t, v_{\infty} t)$; it was computed by applying the
same steps as for the initialization but with the vortex center
translated by the appropriate amount, so that the cell-averaging of
the conserved variables was consistent between the exact solution and
the initial condition.

\begin{figure}
\epsscale{0.9}
\plotone{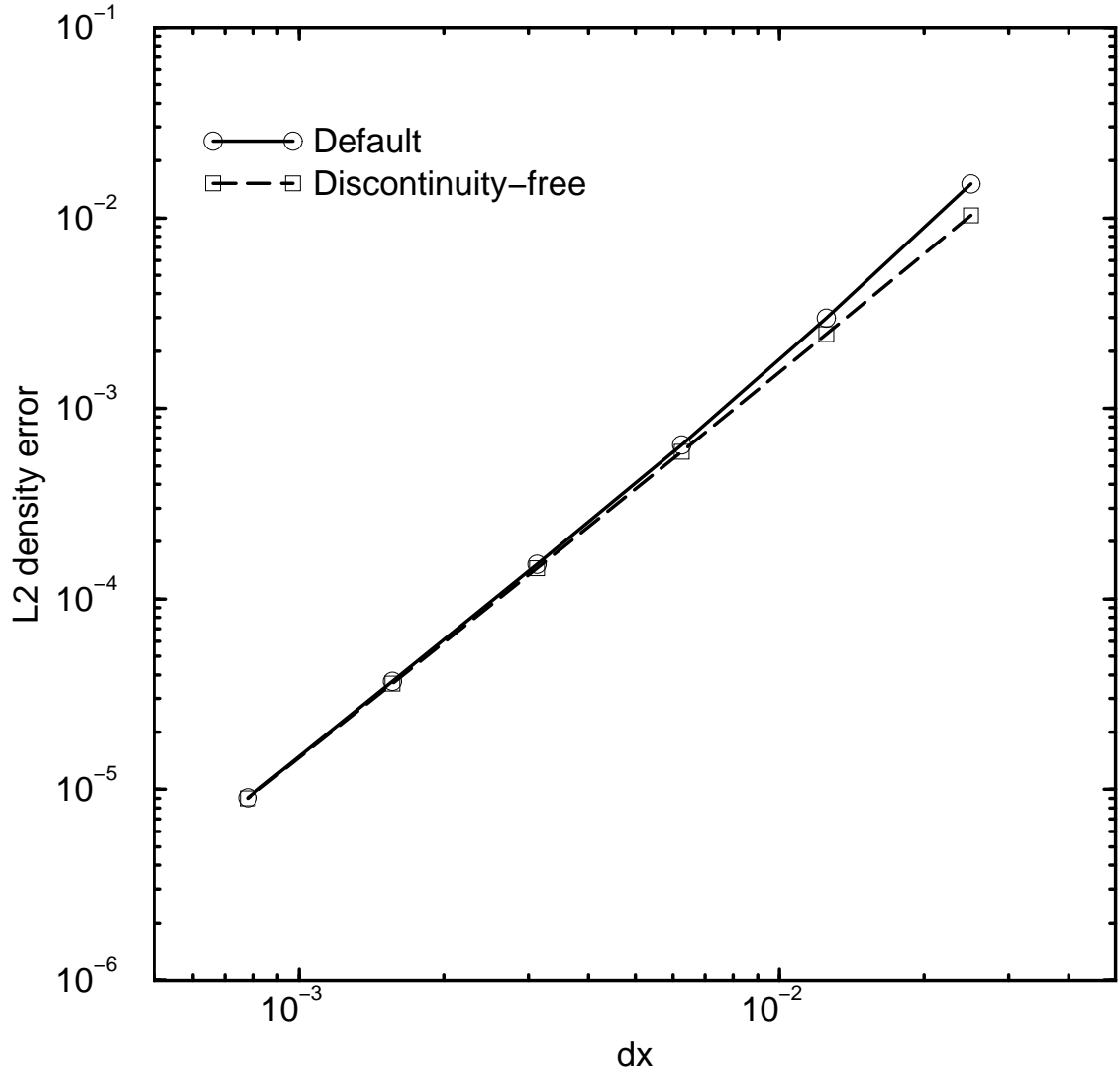}
\epsscale{1.0}
\caption{Plot of L2 density error vs.\ mesh spacing for a two-dimensional advection 
test consisting of an isentropic vortex propagating diagonally across the 
simulation mesh. Shown are results from two sets of simulations, one with full PPM
and one with non-linear steepening turned off. A power-law
fit to the curves gave exponents of 2.13 for the default curve and 2.03 for the 
discontinuity-free curve. 
\label {fig:isenvor}
}
\end{figure}

Figure~\ref{fig:isenvor} shows the $L_2$ error in density is 
plotted vs.\ the mesh spacing for two cases of the isentropic 
vortex problem.  The ``default'' case was computed using the code as it
is most often used for production simulations: no tuning of PPM
parameters or adjustments to the algorithm were made. 
In the ``discontinuity-free'' case, several non-linear components of the 
PPM algorithm that improve behavior at shocks and contacts were
disabled. Contact
steepening, shock flattening, monotonization, and artificial viscosity
are designed to improve or stabilize computations that contain
discontinuous flow features, at the possible expense of increasing
local truncation errors and reducing the convergence rate. Since the 
solution of the isentropic vortex problem is smooth these components 
are not required, and by comparing the discontinuity-free case to the default case,
their influence can be examined. 

Figure~\ref{fig:isenvor} shows that the error decreases as expected
as the mesh is refined; power law fits give convergence rates
of 2.13 for the default simulation and 2.03 for the discontinuity-free simulation.
The error is slightly higher for the default case on coarser
grids, but on the finer grids the default and discontinuity-free versions of the code
produce essentially identical results.
We have varied many aspects of these simulation to see their
effects on the error and the convergence rate. We found that while
the measured errors varied, the convergence
rates were not sensitive to the details of the initialization process
(point-values, interpolation, number of subintervals for estimating
cell-averages), to smaller time steps (and Courant numbers),
or to whether or not the vortex was advecting. In all cases the code
demonstrated second-order convergence.

The verification tests presented here address only the hydrodynamics
module. Great care must be taken in the operator splitting when coupling 
the hydrodynamics module to other modules such as gravity so as to
not degrade the time accuracy. Coupling the hydrodynamics module
to non-time-centered body forces, for example, will produce
first-order convergence in time despite the expected second-order 
convergence of PPM. Additional verification tests are being performed 
with other modules coupled to the hydrodynamics and with a modified version
of the PPM hydrodynamics module for maintaining hydrostatic 
equilibrium~\citep{zingale02a}.

\subsection{Equation of State}

Accurate modeling of astrophysical processes requires 
incorporating as much of the relevant physics as possible.
Realistic models typically require the use of 
physically-motivated equations of state that may be 
experimentally known only over a limited range of conditions
or poorly understood
theoretically. A simulation may naturally
wander into regimes that are not completely covered by these equations
of state or into regions where the equation of state may not be 
thermodynamically consistent. In response, we developed a 
three-part test suite that each
equation of state must pass before we consider using it in a
simulation.  In each part of the suite, multiple calls to the
equation of state using forward (internal energy as a function of
density and temperature) and backward (temperature as a function of
density and internal energy) relations for a given chemical
composition are used to assess the consistency of the equation
of state in
a pre-defined region of thermodynamic variables (density, 
temperature, and chemical composition).
The first test is a uniform scan of the pre-defined region
testing the consistency of the forward and backward calls at a 
set of points uniformly spaced in a logarithmic scale. 
The second test is a random scan of the pre-defined region.
Both tests check the consistency
of the equation
of state by comparing the initial temperature used for
the forward call to that obtained by the backward call.
The third test checks the consistency of the equation of state in the context
of the hydrodynamics method. 
In this case, two hydrodynamic 
states are chosen randomly, and the accuracy
of the solution to the corresponding Riemann problem is
recorded after a fixed number of iterations. If the equation of state is 
inconsistent, it may be impossible to obtain a converged solution.
In general, the accuracy of the equation of state should not 
be worse than the numerical accuracy of the hydrodynamic module 
(typically one part in $10^4$). For verification tests, we require an 
accuracy of 1 part in $10^6$.

\begin{figure}
\plotone{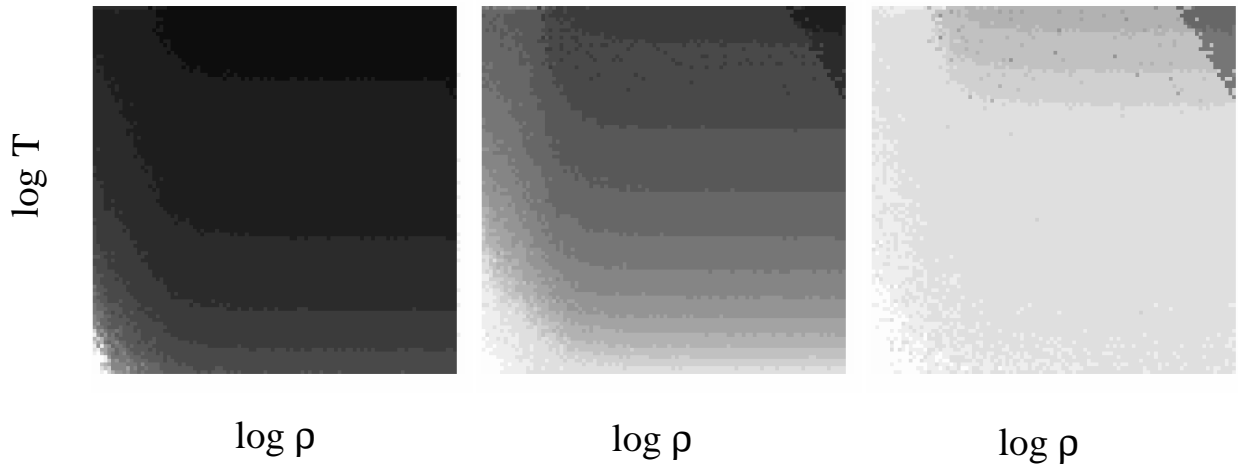}
\caption{Results of testing the electron/positron equation of state. 
The panels from left to right show the results with 5, 7, and 9 Riemann 
solver iterations. Each panel represents the density-temperature plane in log-log
scale, with density on the x-axis and temperature on the y-axis.  
The density ranged from $10^{-30}$ to $10^{-19}\gcc$ and 
the temperature ranged from $10^{4}$ to $10^{11}$ K.
The gray scale indicates a relative error between 
$10^{-17}$ (white) and $10^{-1}$ (black).
\label {fig:eostest}
}
\end{figure}

To demonstrate this procedure, we applied the equation of state test suite to 
an electron-positron equation of state \citep{muller01} that would be applicable 
to a high-temperature ($T \geq 10^9$ K) plasma that might occur in an 
astrophysical environment such as the vicinity of a pulsar. In its current 
implementation, the equation of state does not depend on the chemical composition. 
The equation of state was tested in the region of interest defined as 
$10^{-30} < \varrho < 10^{-19}\gcc$  and $10^{4} < T < 10^{11}$ K. 
We selected $10^7$ ($\log_{10} \varrho, \log_{10} T$) pairs, and binned the
results into a 100 by 100 point ($\rho - T$) array. The results are presented 
in Figure~\ref{fig:eostest}, with the gray scale
indicating the relative error for the Riemann solver with this EOS 
between $10^{-17}$ and $10^{-1}$ for 5 (left panel),
7 (middle panel), and 9 (right panel) Riemann solver iterations.
The equation of state is the most accurate in the low density, low temperature
regime (lower left corners of the panels), and its accuracy  decreases gradually
as the density and temperature increases. We note that even for 9 Riemann
solver iterations, the largest relative error on the domain is never less than $10^{-6}$.
We attribute this limit to the accuracy of the 
approximations of the Bessel functions used in the calculations.
The fraction of cases for which the relative error exceeds $10^{-6}$ is 
49, 6, and 0\% for 5, 7, and 9 iterations, respectively.
Table~\ref{tbl:1} presents the distribution of error for 5 
Riemann solver iterations (left panel in Figure~\ref{fig:eostest}).
The results of this study led us to the conclusion that eight iterations
of the Riemann solver is sufficient to achieve the desired accuracy
for simulations using this particular equation of state.
In general, though, we perform simulations with a test of
convergence to a relative error rather than a fixed number of 
iterations.

In modeling the three-layer target experiments (see section below), we began
with a modified version of Sesame equation of state tables \citep{sesame}. 
In the original form, the Sesame table is not suitable for use in 
conservative hydrodynamic simulations because it only provides pressure 
and energy as a function of density and temperature (forward relation),
while a conservative simulation requires pressure and temperature as a 
function of density and energy (backward relation). 
The latter relation in principle can be obtained 
by a numerical inversion of the table, but the relation must satisfy thermodynamic 
consistency and accuracy desired by the hydrodynamic module.
The modified tables we tested included inverted tables, but 
our prescription for testing an equation of state showed that the modified Sesame tables
did not satisfy our criteria for use in a validation problem. 
Finding and testing other available equations of state, e.g. QEOS \citep{more88},
are the subjects of ongoing research.

\section{Validation Tests}

In the following sections, we present the results of validating FLASH with
two laboratory experiments. These efforts focus only
on validating the principal hydrodynamics module in FLASH, and the validation
of other code modules such as burning and gravity is the subject of ongoing
research. Where possible, we have quantified the results of the 
simulation-experiment comparison. We note again, however, that
an important part of validation is finding an acceptable
physically-motivated equation of state appropriate for the materials in 
the experiment. As described above, we found that testing the equation of state in the
context of the hydrodynamics method is important. Otherwise, a problem that
is meant to validate a code by comparison to experiment becomes instead
a lengthy analysis and validation (or not) of a model equation of state.
Because the intent of this work was validation of the hydrodynamics module in
FLASH and because the difficult problem of validating a material equation 
of state for terrestrial materials is beyond the scope of our efforts,
the simulations presented below made use of simple gamma-law equations of 
state.

The validation tests presented below were performed on an adaptive mesh. In the case of
the laser-driven shock simulations, the standard criteria for mesh refinement
(testing the magnitude of the second derivative of density and pressure and refining or
de-refining the mesh in regions where the magnitude is above or below a threshold) 
worked well to capture the shocks and discontinuities of the flow. In the 
Rayleigh-Taylor simulations, to avoid the possibility of under-resolving the initial 
conditions, the simulation domain in the region of the initial perturbations was 
forced to be fully refined. Beyond that region, the simulation applied the 
standard criteria for mesh refinement.

\subsection{Laser-driven Shock Simulations}

Intense lasers offer the chance to probe experimentally
environments similar to those that exist in complex astrophysical 
phenomena. Such experiments are obvious choices for code validation.
\citet{holmes99} performed a careful study of such experiments,
investigating the  Richtmyer-Meshkov~\citep{richtmyer60,meshkov69} instability 
for negative Atwood numbers and 
two-dimensional sinusoidal perturbations. This study included experimental, 
numerical, and theoretical work and produced a quantitative comparison between 
results. Our efforts focus on modeling experiments performed using the 
Omega laser facility at the University of Rochester 
\citep{soures96,boehly95,bradley98} 
that involve shock propagation through a multi-layer target. These experiments are 
designed to replicate the 
hydrodynamic instabilities thought to arise during supernova explosions.
In addition to validation, the experiments may provide a 
better understanding of the turbulent mixing that occurs as a result of
instabilities driven by the propagation of a shock through a layered target. 

The experiment we used for validation consists of a strong
shock driven through a target with three layers of decreasing density. 
The interface between the first two layers
is rippled while the second interface is flat. The planar shock 
is perturbed as it crosses the first interface and excites a Richtmyer-Meshkov
instability. The perturbed shock then propagates through the second interface,
imprinting the perturbation on the interface and leading to the 
growth of additional fluid instabilities.
This three-layer experiment is meant to model the configuration of a 
core collapse supernova. In this case, it has been proposed that the 
development of fluid instabilities followed by mixing is responsible for 
certain features present in spectra obtained
during the first few hundred days after the explosion~\citep[e.g.][]{arnett89}.
The accepted scenario involves a supernova shock propagating
through the outer layers of the star, which is composed of shells
of different chemical compositions. The interaction of the shock with
the interfaces between these shells leads to the development and growth of 
Richtmyer-Meshkov and Rayleigh-Taylor instabilities. 
These instabilities can grow from seed perturbations provided
at the shock front by convection 
inside the proto-neutron star~\citep{burrows95,keil96,mezz98a} and/or by 
neutrino-driven convection behind the 
shock~\citep{miller93,herant94,janka96,burrows95,mezz98b}. 
Other seed perturbations can arise by convective burning prior to 
collapse~\citep{bazan98,heger00}.
The effect of such instabilities would be mixing of the material in the core
of the star with material in the outer regions. This process may be able
to explain the early observation of radioactive core elements 
in SN 1987A \citep[and references therein]{kifonidis00}.

The target consists of three main layers of material in a cylindrical Be
shock tube, with the initial density decreasing in the direction of
shock propagation. The materials are Cu, polyimide plastic, and 
carbonized resorcinol formaldehyde (CRF) foam, with thicknesses of 85, 
150, and 1500$\microm$ and densities 8.93, 1.41, and 0.1$\gcc$, 
respectively. Performing the experiment with the target inside a shock 
tube delays the lateral decompression of the target, giving a more planar 
shock. Be is chosen as the material for the shock tube
as it is essentially transparent to the diagnostic X-rays. The surface of 
the Cu layer is machined 
with a sinusoidal ripple of wavelength $200\microm$ and amplitude $15\microm$. 
The laser drive end of the target consists of a $10\microm$ section of CH ablator to 
prevent direct illumination of the target and the associated pre-heating of the 
rest of the target.  Embedded within the polyimide 
layer is a $75\microm$ thick,
$200\microm$ wide (along the diagnostic line of sight)
tracer strip of brominated CH (4.3\% by number of atoms in the material, i.e. the atomic 
composition is C$_{500}$H$_{457}$Br$_{43}$). 

The experiment is driven by 10 beams of the Omega laser with a nominal 
measured energy of 420 J/beam in a 1 ns pulse at a laser wavelength of 
$\lambda_L = 0.351\microm$. The peak intensity (in the overlapped spot) is 
$7.2 \times 10^{14}\wpcm$, while the average intensity is 
$5.7 \times 10^{14}\wpcm$. The shock is perturbed as 
it crosses the corrugated Cu-polyimide interface and oscillates as it 
propagates through the polyimide/CH(Br). When it reaches the foam interface, 
it imprints the perturbation. The experiment is observed side-on with 
hard X-ray radiography using a gated framing camera. Eight additional laser 
beams are focused on an iron back-lighter foil located near the target and 
generate 6.7 keV X-rays to which the Cu and CH(Br) tracer 
strip are opaque and the polyimide and
foam are nearly transparent. Nearly all of the contrast at the 
polyimide/CH(Br)-foam interface comes from the tracer layer. This allows
visualization of the shock-imprinted structure at that interface over only 
the central $200\microm$ of the target along the line of sight without 
edge effects near the wall of the shock tube.
Full details of the experiment may be found in \citet{kane01} and \citet{robey01}.

A recent study by \citet{robey02} addressing the issue of the onset of
turbulence in laser-driven shock
experiments provides an estimate of the Reynolds numbers 
for these experiments.  The study notes that
recent experimental work by~\citet{dimotakis00} indicates that for a
wide range of stationary flow geometries, there appears to be a nearly
universal value of Reynolds number (1-2 $\times 10^{4}$) at which an
abrupt transition to a well-mixed state occurs, and this transition has
been suggested as an indicator for the transition to fully-developed
turbulence. By combining experimental measurements with a kinematic
viscosity estimate, the Robey et al.\/ study indicates that the time
dependent Reynolds number can approach $10^{5}$ during the laser-driven
shock experiments and that though there are some caveats, the
experiments are ``perhaps very close but somewhat short of the
threshold value required for the onset of a mixing transition." This
would indicate that these sorts of experiments may be approaching the
transition to fully-developed turbulence.  We note that the
laser-driven experiment described in this study differs somewhat from the
laser experiment of our validation test, but it demonstrates the flow
regimes that laser-driven shock experiments can reach.

\begin{figure}
\plotone{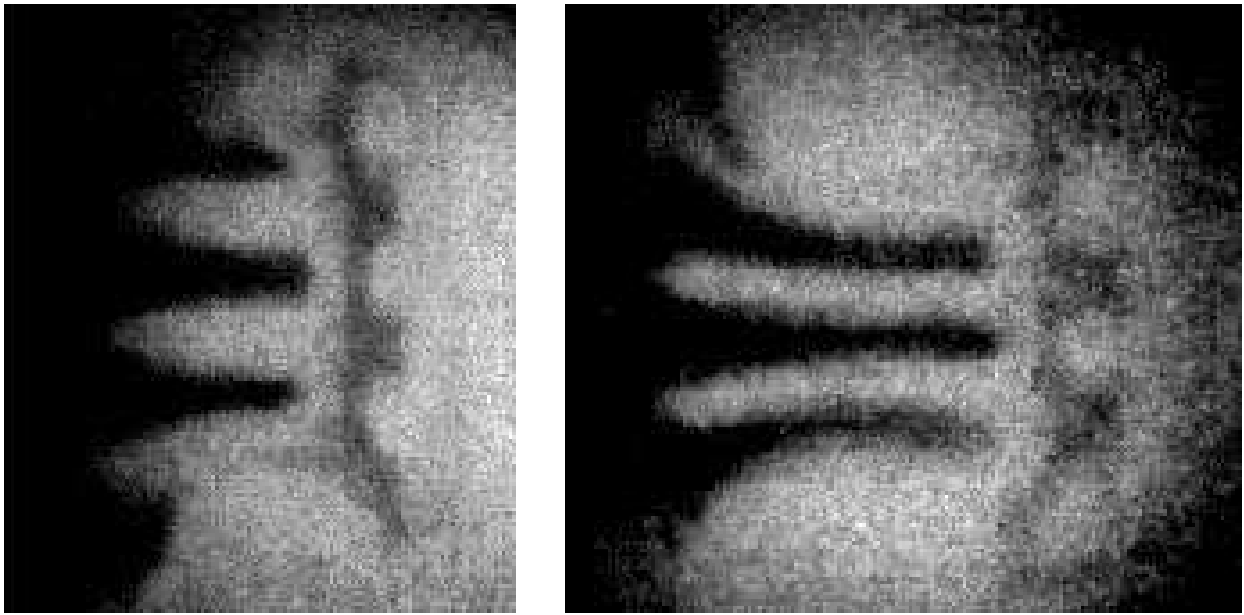}
\caption{Results of the 3-layer target experiment. Shown are side-on X-ray radiographs at
39.9 ns (left) and 66.0 ns (right). The long, dark ``fingers" are spikes of expanding 
Cu, and the horizontal band of opaque material to the right of the spikes of Cu
is the brominated plastic tracer showing the imprinted instability growth
at the plastic-foam interface.
\label{fig:3lay_exp}
}
\end{figure}

Figure \ref{fig:3lay_exp} shows the results of the 3-layer target 
experiment. The images are X-ray radiographs at two times, 39.9 ns (left) and 
66.0 ns (right). The long, dark ``fingers" are spikes of expanding
Cu, and the horizontal band of opaque material to the right of the spikes of 
Cu is the brominated plastic tracer, showing the imprinted instability growth
at the plastic-foam interface.
The length of the Cu spikes in the experiment was determined by three
methods.  The first method was a straightforward visual inspection of the
images using as a spatial reference a gold grid of $63.5\microm$ period, located
just below the images of Figure \ref{fig:3lay_exp}. The second method used a contour
routine to try to better quantify the uncertainty in the location of the
edges of the spikes.  The third method was done in a manner consistent
with the analysis of the numerical simulations.  A $500\microm$ section in the
center of the images was vertically averaged to produce a single spatial
lineout of optical depth through the region occupied by the Cu and CH.
The same 5\% and 90\% threshold values were used to quantitatively determine
the extent of the Cu spikes.  Taking the average of all three methods,
values of 330 $\pm$ $25\microm$ and 554 $\pm$ $25\microm$ are obtained 
at 39.9 and 66.0 ns, respectively.

There are several sources contributing to error in these experimental
measurements: the spatial resolution of the diagnostic, the
photon statistics of the image, target alignment and parallax, and the
specific contrast level chosen as the definition of the length of the
Cu spikes.  The intrinsic diagnostic resolution is set by the imaging
pinhole, which is $20 \microm$ in diameter.  The photon noise statistics of
the image can further degrade the spatial resolution when the number of
photons per detector resolution element is small \citep[see recent reference:][]{landen01}.
For the present number of backlighter beams,
diagnostic magnification, and pinhole size, the photon statistics produce
a signal-to-noise ratio $>$ 20, and therefore do not contribute to a further
decrease in spatial resolution.  The target alignment with respect
to the diagnostic line-of-sight is generally within 1\%.  Since most
of the contrast comes from the $200\microm$ wide radiographic CH(Br) tracer
layer, this contributes an additional 3.5\% to the spatial uncertainty.
Finally, the specific contrast level chosen to represent the spatial
extent of the spikes as discussed above contributes to the uncertainty.
This uncertainty was estimated from the variation resulting from the
three methods used to determine the spike lengths.

In addition to the spatial error, there are also several sources of
uncertainty in the temporal accuracy of the measurements.  These arise
from target-to-target dimensional variations, shot-to-shot drive intensity
variations, and the intrinsic timing accuracy of the diagnostics.
The nominal target dimensions used in the simulations were $10\microm$ of CH
(ablator layer), $85\microm$ Cu, $150\microm$ of polyimide CH(4.3\% Br), and $1500\microm$
of CRF.  In the actual targets, however, there is some variation from
these nominal values, and this variation enters into the quantification of temporal
error between experiment and simulations.  For the target that produced
the image at 39.9 ns, the dimensions of the 4 layers are 10, 105, 145,
and $1505\microm$.   For the target that produced the image at 66.0 ns, the
dimensions of the 4 layers are 10, 98, 150, and $1514\microm$.  The biggest
difference between the experimental target dimensions and those used in
the simulations is in the dense Cu layer, where the experimental targets
are $\sim 20$\% thicker.  Shock propagation through these slightly different
thicknesses will cause a small discrepancy in the timing.

The variation in the drive intensity from shot-to-shot is another possible
source of temporal uncertainty.  For the image obtained at 39.9 ns, the
average laser drive energy was 416 J(UV)/beam.  For the shot at 66.0 ns,
the drive energy was 421.5 J(UV)/beam.  Therefore, there is a 1.3\%
intensity variation.  The drive pressure scales as the intensity to
the 2/3 power, and material velocities scale as the square root of the
pressure \citep{lindl98}, so the velocity variations from shot-to-shot
will be less than 0.5\%.  Shot-to-shot timing variations resulting from
this mechanism will be of the same order, i.e. less than one percent.
The intrinsic temporal accuracy of the diagnostics is even better
than this, with a typical uncertainty of $\pm$ 200 ps.  The experimental
uncertainty in the timing is therefore relatively small, and is approximately
indicated by the width of the symbols used in the figure (described below)
that compares the experimental results to the simulation results.

\begin{figure}
\plotone{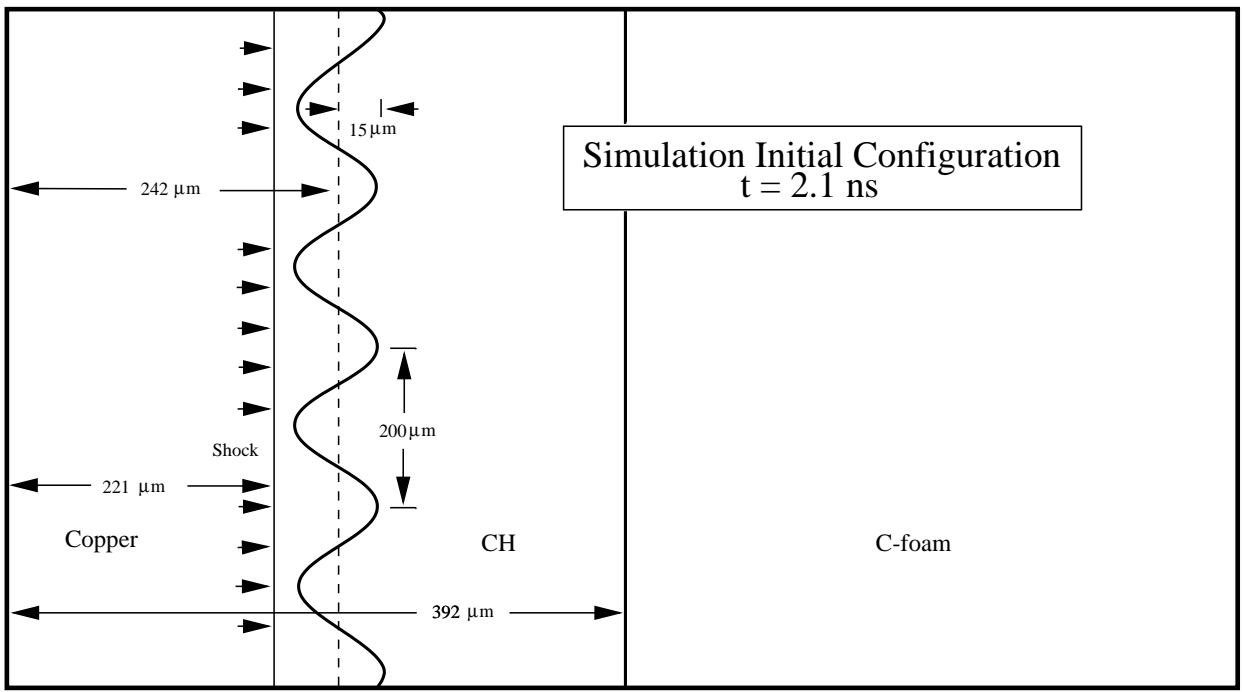}
\caption{Schematic of the 3-layer target simulation initial conditions.
Shown are the locations of the three materials, Cu, CH,
and C, the shock, and the details of the sinusoidal perturbation
of the Cu-CH interface. The schematic is not to scale.
\label{fig:3lay_scheme}
}
\end{figure}

The simulation models the experiment with a similar three-layer arrangement.
The three materials were Cu, polyimide CH, and C with the 
same densities as the three layers of the actual target. The simulation
began 2.1 ns into the experiment, at which point the shock is approaching
the Cu-CH interface. The initial thermodynamic profiles were obtained
from simulations of the laser-material interaction performed with a 
one-dimensional radiation hydrodynamics code~\citep{larsen94}. 
The results were mapped onto the two-dimensional grid with a perturbed 
Cu-CH interface at a simulation time of 2.1 ns and were then evolved out to 
approximately 66 ns. The materials were modeled as gamma-law gases, with 
$\gamma =$ 2.0, 2.0, and 1.3 for the Cu, CH, C, 
respectively. These values for gamma were chosen to give 
similar shock speeds to those observed in the experiments. 
Figure \ref{fig:3lay_scheme} illustrates the initial configuration. The simulation 
used periodic boundary conditions on the transverse boundaries and zero-gradient 
outflow boundary conditions on the boundaries in the direction of the shock 
propagation. Note that use of periodic boundary conditions in
the transverse directions is not in keeping with the boundary conditions of
the experiment. The experiment was performed with the three materials of the
target inside a cylindrical Be shock tube. Accordingly, the experiment results
show the influence of the shock tube walls as a curving or pinching of the outer
Cu spikes. Our simplified model did not consider these boundary effects.

\begin{figure}
\plotone{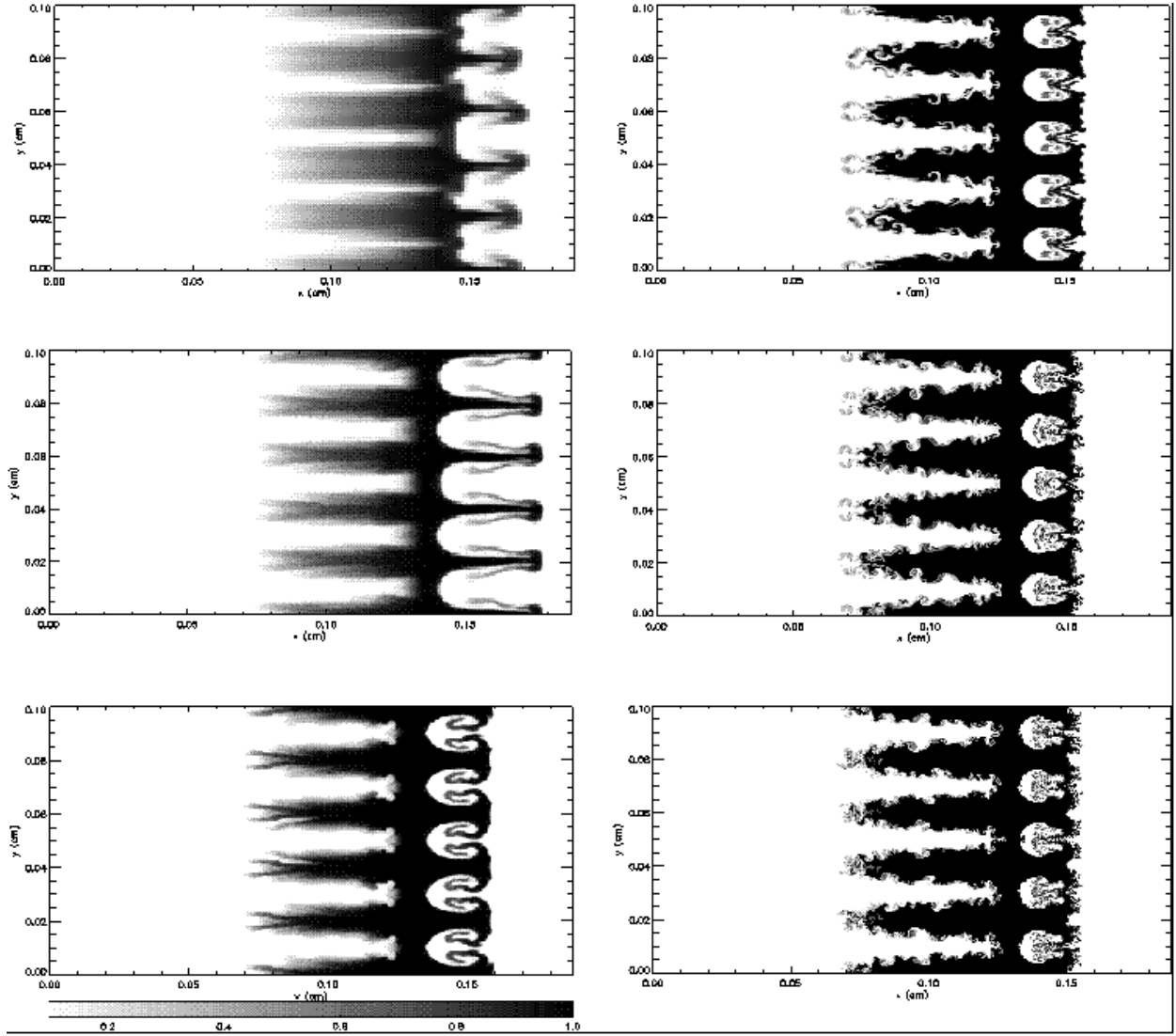}
\caption{Gray scale images of CH abundance at approximately the time of the
late time experimental image, 66.0 ns, from simulations
at varying resolutions. The effective simulation resolutions were, top to bottom
on left followed by top to bottom on right, 128 $\times$ 64, 256 $\times$ 512,  
512 $\times$ 1024, 1024 $\times$ 2048,  2048 $\times$ 4096, corresponding
to 4, 5, 6, 7, 8, and 9 levels of adaptive mesh refinement.
\label{fig:6ch}
}
\end{figure}

We first present the results of a resolution study of the simulation.
We performed six simulations from equivalent initial conditions, changing only 
the maximum mesh resolution. The results of the 
six simulations are illustrated in Figure \ref{fig:6ch}. 
As described above, FLASH solves an advection equation for each abundance,
allowing us to track the flow of each material with time.
Shown are fluid abundances for the CH, the intermediate material, at approximately 66 ns. 
The abundance is represented by a gray scale, so that the white regions 
(for which the abundance is zero) to the left and right of the central gray region 
are Cu and C, respectively. Thus, the CH abundance shows
features of both material interfaces, and its evolution shows the 
behavior of both the spikes of Cu and bubbles of C. 
The effective simulation resolutions were, top to bottom on left then top to 
bottom on right, 128 $\times$ 64, 256 $\times$ 512,
512 $\times$ 1024, 1024 $\times$ 2048,  2048 $\times$ 4096, corresponding
to 4, 5, 6, 7, 8, and 9 levels of adaptive mesh refinement. 

We found that all of the simulations capture the expected bulk properties of the 
flow; that is, all showed similar spikes of Cu and bubbles of C. The principal 
differences in the results were in the amount of small scale structure in the flow, 
with the amount of small-scale structure increasing with resolution. 
The increasing amount of small scale structure in the simulations should be 
readily apparent in the panels of the figure. This behavior is
expected because the dissipation mechanism in PPM operates on smaller and smaller
scales as the resolution is increased. 
We note that the increasing amount of small-scale structure seen
with increasing resolution indicates that these simulations will
not demonstrate a converged flow with higher resolution until the
turbulent scales are fully resolved; only at this point will we
be resolving all the relevant length scales in the problem. 

Because, as noted above, we observe increasing amounts of small-scale structure
in the simulations with higher resolutions,
convergence studies should focus on integrated quantities such instability
growth, which may converge.
In order to quantify the results, we examined the lengths of the Cu spikes,
an integral quantity of the flow that we were best able to accurately measure both 
in the simulation results and in the experimental results.
The spike lengths were measured by averaging the CH abundance
in the $y$-direction across the simulation domain then smoothing the resulting
one-dimensional array slightly to minimize differences that would occur
owing to very small scale structure. The length of the Cu spikes was then 
determined by the average distance spanned by minimum locations of average abundances 
0.05 and  0.9. The results were reasonably robust to the amount of smoothing
and threshold values.

\begin{figure}
\plotone{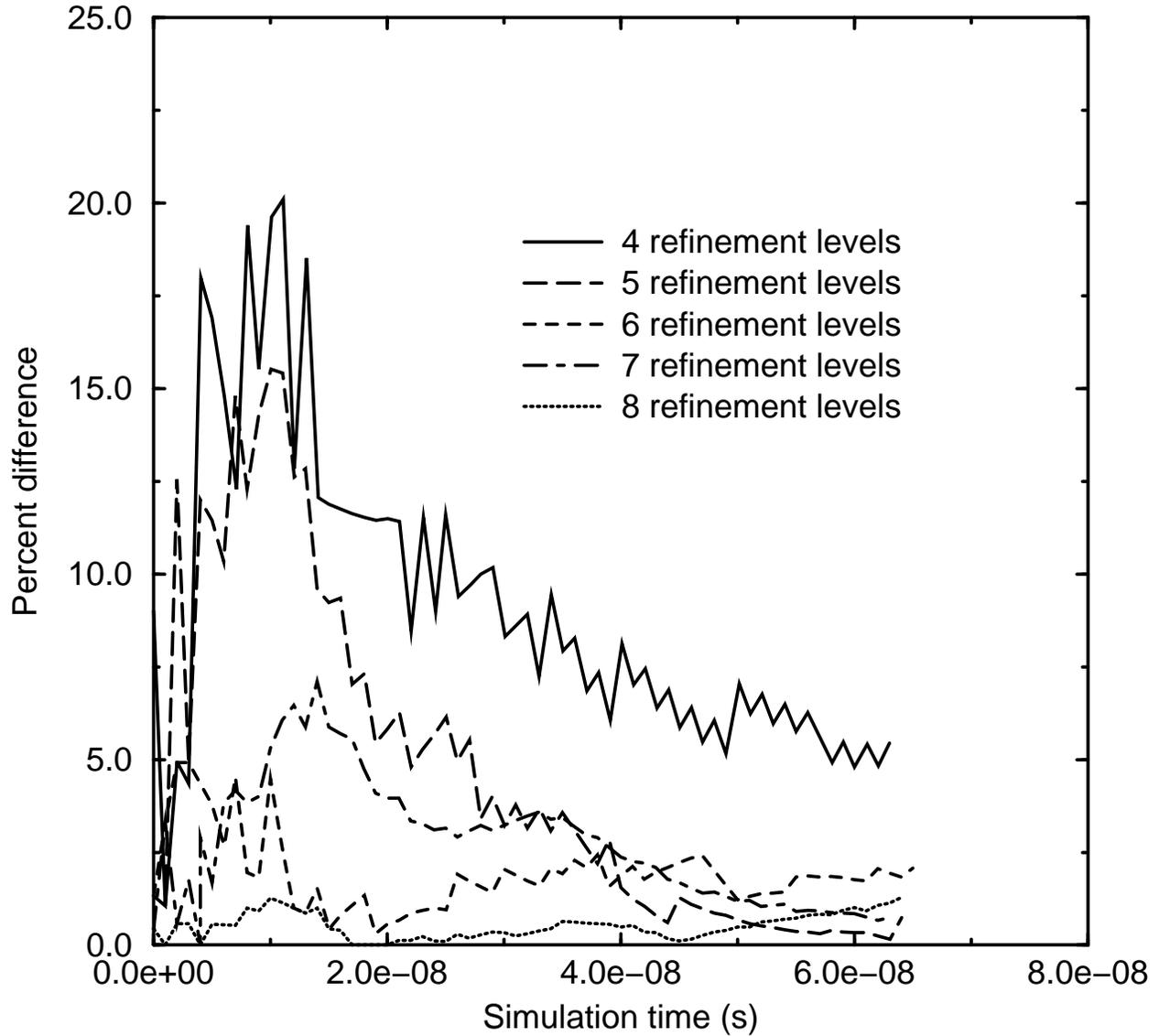}
\caption{Percent difference of the Cu spike lengths from those of 
the highest resolution (9 levels of adaptive mesh refinement)
simulation vs.\ time. The percent differences are from the lower resolution
simulations of 4, 5, 6, 7, and 8 levels of adaptive mesh refinement.
We note that the convergence is not perfect. The curve from the 8 level of refinement
simulation crosses those of the 6 and 7 level of refinement simulations, indicating
a higher percent difference.
\label{fig:percentd}
}
\end{figure}

\begin{figure}
\plotone{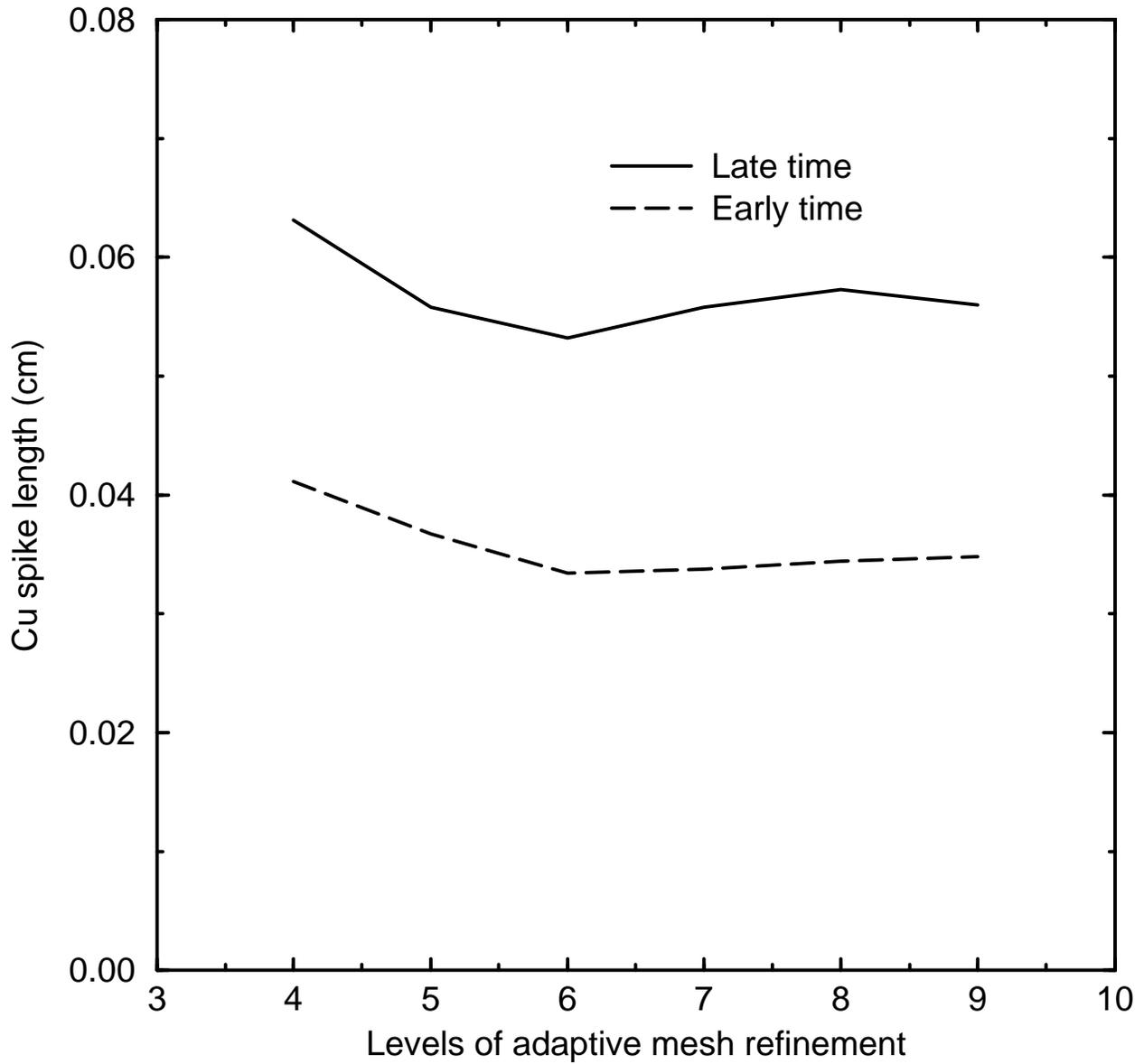}
\caption{Cu spike length vs.\ adaptive mesh refinement level. Shown are the spike
lengths at the two times corresponding to those of the experimental results
from all of the simulations. The percent differences shown in Figure \ref{fig:percentd}
were calculated at many times in the simulations. This figure illustrates how
the spike length changed with resolution at the two times of the experimental results. 
\label{fig:spikes_2times}
}
\end{figure}

The results of testing the convergence of the Cu spike length measurements
are shown in Figures \ref{fig:percentd} and \ref{fig:spikes_2times}. 
Figure \ref{fig:percentd} shows percent differences
from the highest resolution simulation as functions of time. The trend is
that the difference decreases with increasing mesh resolution, with 
the seven and eight level of adaptive mesh refinement simulations always 
demonstrating agreement to within five percent. The trend of decreasing 
difference with increasing mesh resolution demonstrates a convergence 
of the flow, but it is subject to caveats. We note that the trend does not 
describe the behavior at all points in time (that is, the percent difference 
curves sometimes cross each other), and this average measurement is an 
integral property of the flow and in no way quantifies the differences in 
small scale structure observed in the abundances. In particular,
we note that the difference curve for the simulation with eight adaptive
mesh refinement levels
crosses the curves of both the seven and six level simulations, suggesting
that higher resolution simulations may deviate further from these results
and produce degraded agreement with the experiment.
Figure~\ref{fig:spikes_2times} shows the Cu spike lengths at approximately the 
two times of the experimental results, 39.9 and 66.0 ns, vs.\ number of 
refinement levels, quantifying the changes of the Cu spike lengths at the 
times of the experimental results with resolution.

\begin{figure}
\plotone{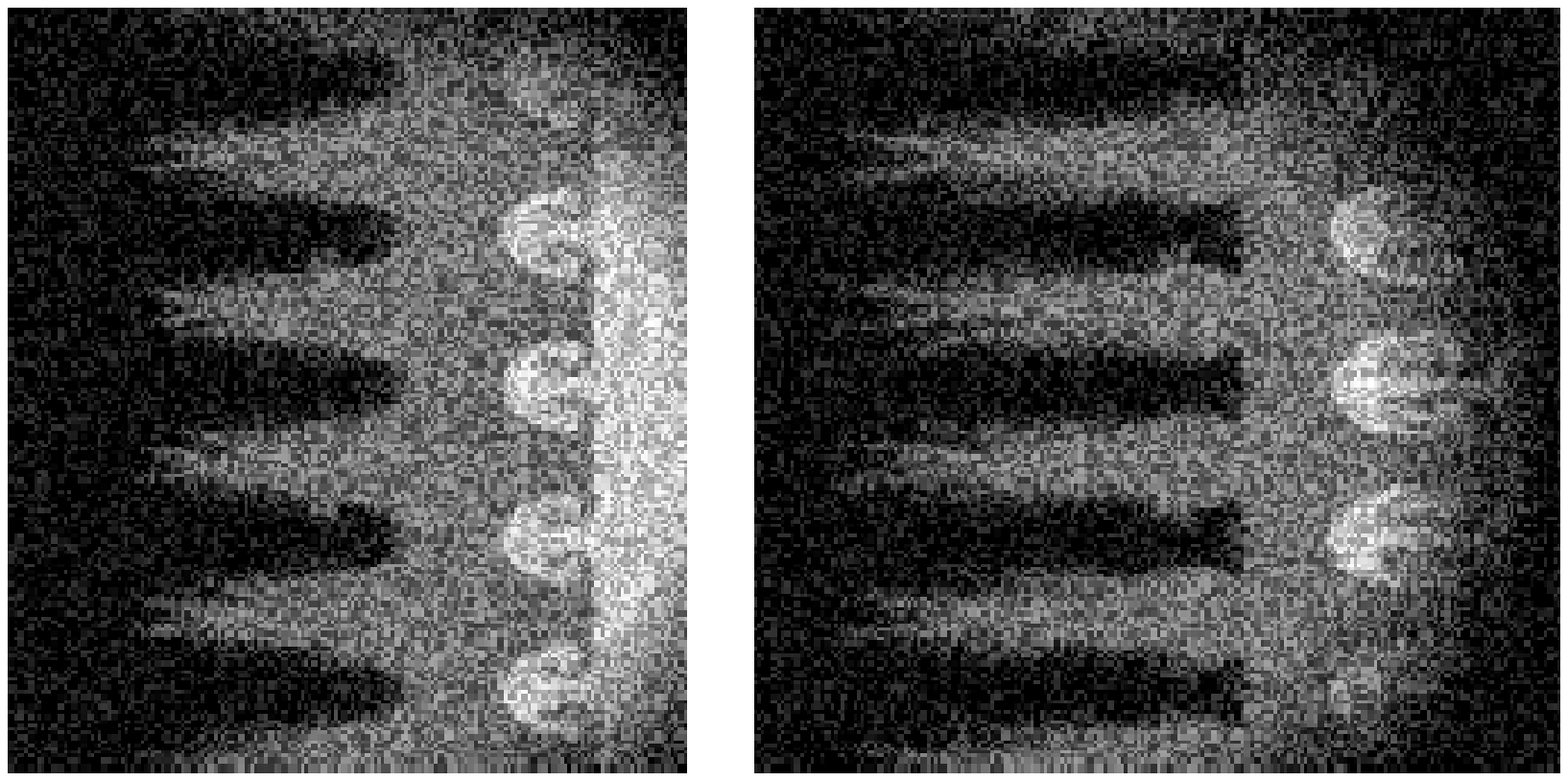}
\caption{
Simulated radiographs from the six level of refinement 
(effective resolution of 512 $\times$ 256) simulation of the three-layer
target experiment. The simulated radiographs were created from the fluid 
abundances at times corresponding approximately to those of the images from 
the experiment, 39.9 ns (right) and 66.0 ns (right). 
Shown are the parts of the simulation domain that
match the regions in the experimental results.
\label{fig:simrad6}
}
\end{figure}

\begin{figure}
\plotone{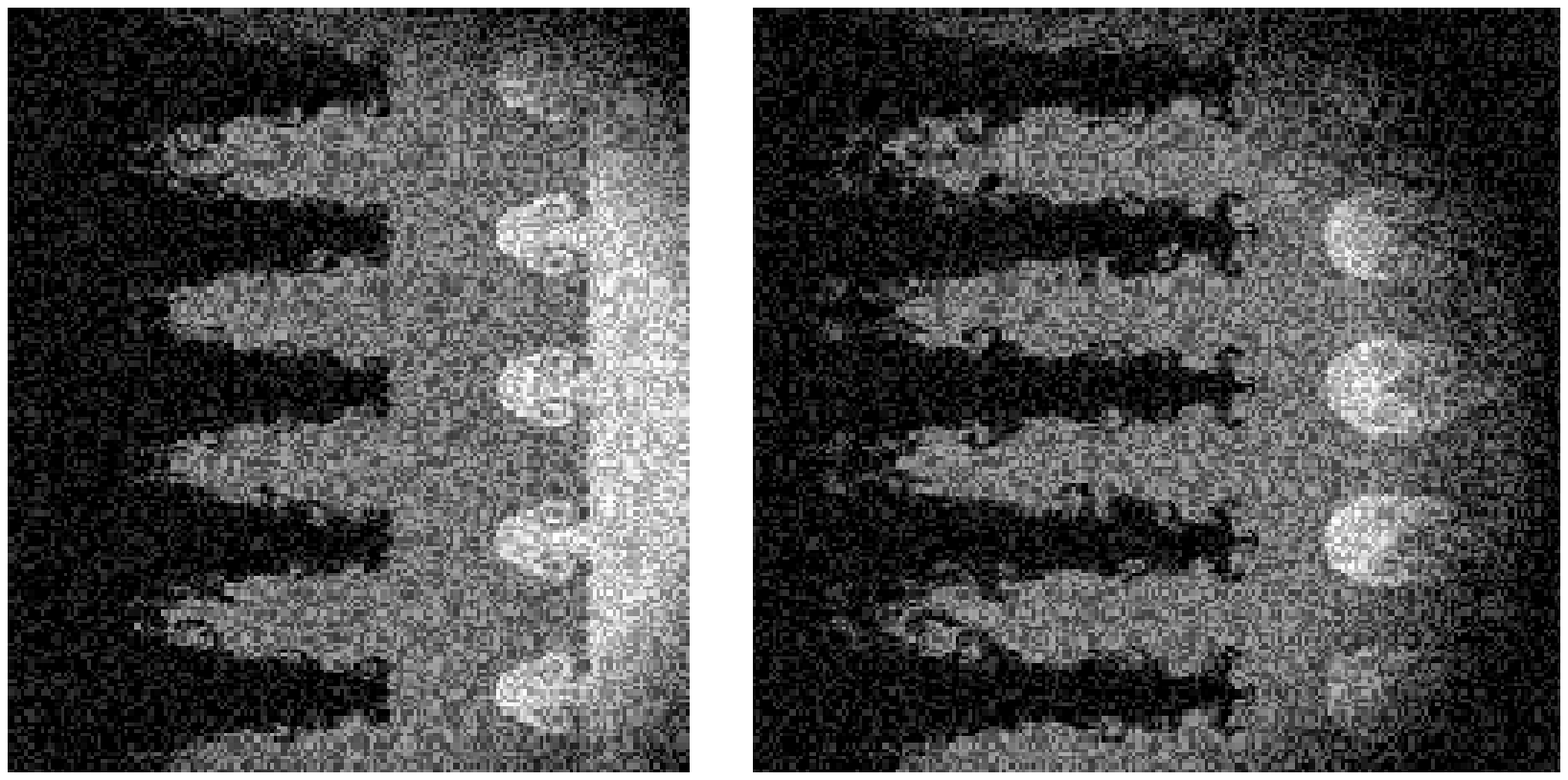}
\caption{
Simulated radiographs from the seven level of refinement 
(effective resolution of 1024 $\times$ 512) simulation of the three-layer
target experiment. The simulated radiographs were created from the fluid 
abundances at times corresponding approximately to those of the images from the experiment, 
39.9 ns (left) and 66.0 ns (right). 
Shown are the parts of the simulation domain that
match the regions in the experimental results.
\label{fig:simrad7}
}
\end{figure}

Figures \ref{fig:simrad6} and \ref{fig:simrad7} show simulated radiographs 
from two simulations. Both figures show a simulated
radiograph at approximately the two times corresponding to the images from
the experiment, 39.9 ns (left panel) and 66.0 ns (right panel). 
The simulation in Figure \ref{fig:simrad6} had six levels of adaptive mesh 
refinement (an effective resolution of $512 \times 256$), while 
the simulation in Figure \ref{fig:simrad7} had seven levels of 
refinement ($1024 \times 512$).
The radiographs were created from the abundances of the three materials.
An artificial opacity was assigned to each 
abundance, Cu having the highest, CH the intermediate, and C
the lowest. Then the opacity was applied to intensity from an artificial ``beam" that
consisted of a circular region that linearly decayed above a certain
radius from a point near the center of the simulation domain. In addition,
the abundances were de-resolved to match the resolution of the pixels
in the images from the experiment and random Poisson `noise' was added to the 
intensity.  Comparison of the simulated radiographs to the radiographs
from the experiment show that the simulations captured the bulk behavior
of the materials, particularly the growth of Cu spikes and the development 
of C bubbles. 

It is worth noting that the 
increased amount of small scale structure in the higher resolution simulation
is visible in comparing the radiographs even though the images have been 
de-resolved. From inspection, the amount of small scale structure in the
six level simulation appears to agree with that observed in the experiment
better than the seven level simulation. This agreement indicates that 
the amount of viscosity in the experiment is higher than the numerical 
viscosity of the higher resolution runs. Caution concerning this 
interpretation is warranted, however, because the experimental radiograph, 
which is essentially the shadow of the material, will 
almost certainly fail to capture the true amount of small scale structure. 
Conclusions regarding the correct amount of small scale structure 
(and the correct Reynolds number of the flow) must await better diagnostic 
resolution of the experiments followed by a careful, quantitative comparison 
with simulation results.

\begin{figure}
\plotone{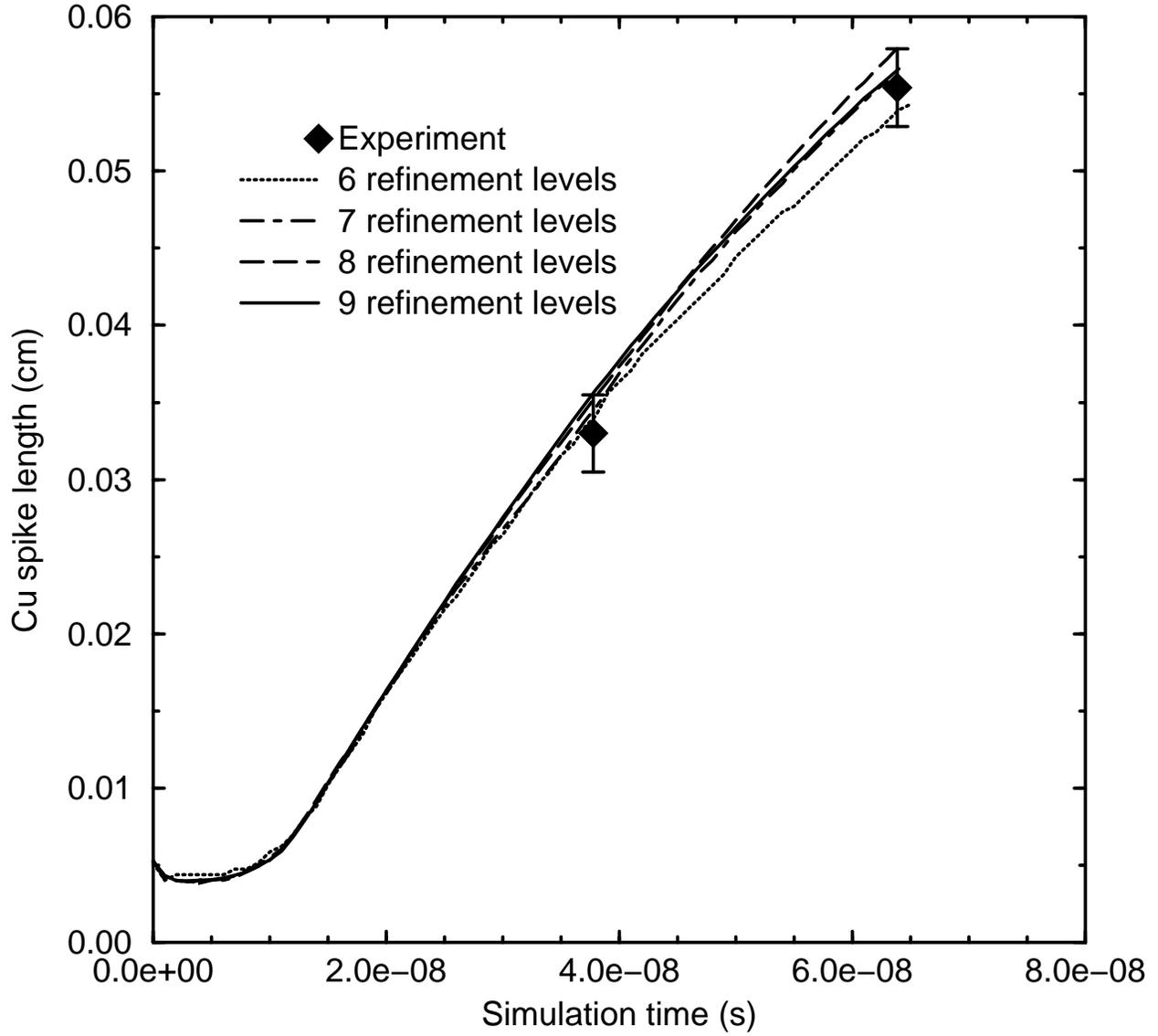}
\caption{
Cu spike length vs. time. The curves are from simulations at 6, 7, 8, and 9 levels of 
refinement simulations (effective resolutions of 256 $\times$ 512, 512 $\times$ 1024, 
1024 $\times$ 2048, 2048 $\times$ 4096), and the points with error bars are results 
from the experiment. The error bars represent $\pm 25\microm$, and the width of 
the symbols represents the timing error.
\label{fig:7andexp4}
}
\end{figure}

\begin{figure}
\epsscale{1.0}
\plotone{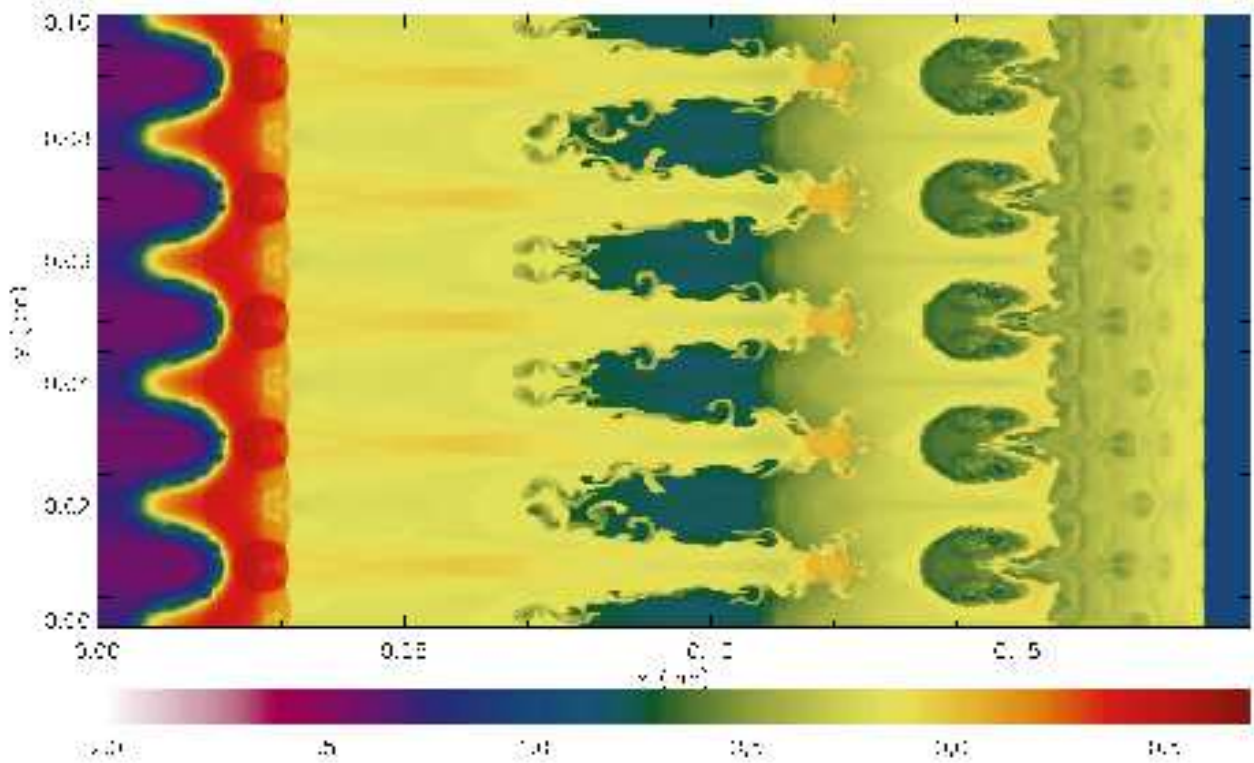}
\epsscale{1.0}
\caption{Full resolution image of the log of density from the 
seven levels of adaptive mesh refinement simulation
at approximately the time of the late time experimental result.
The spikes of Cu are visible as the
reddish-yellow ($\rho \sim 2\gcc$) fingers moving into the 
less dense ($\rho \sim 0.5\gcc$) CH. The bubbles of C
are the dark green ($\rho \sim 0.2\gcc$) regions to the right
of, and opposing, the spikes of Cu. The transition from compressed C
(rightmost yellowish region) to uncompressed C (blue region on far right)
marks the position of the shock, which shows a slight perturbation.
\label{fig:3lay_exp_sim2}
}
\end{figure}

Figure \ref{fig:7andexp4} shows the Cu spike length vs.\ time for the six and seven
levels of adaptive mesh refinement simulations.  Also shown in the figure are 
the experimental results. The error bars correspond to $\pm 25\microm$, the
spatial error of the experiment. The width of the symbols marking the experimental
results indicates approximately the timing error. The 
figure shows that the simulations agree well with the experimental results. 
In addition, inspection of the evolution of the shock as it passes through 
the CH shows that the shock oscillates as expected  from 
the linear theory of compressible fluids \citep{dyakov54,freeman55,ll87} 
and is almost planar when it reaches the CH/C interface.
Figure \ref{fig:3lay_exp_sim2} shows the logarithm of density on the entire 
simulation grid at approximately the time of the late experimental result 
for the simulation with seven levels of adaptive mesh refinement. In this 
case, the image is fully resolved (to the resolution of the simulation).

\subsection{Rayleigh-Taylor Simulations}

The classic Rayleigh-Taylor problem consists of a dense fluid on top of a light
fluid in the presence of a gravitational acceleration. The initial configuration
is in an unstable equilibrium, and any perturbation of the fluid interface
leads to instability growth. In the linear regime, the
instability growth rate is proportional to the
square root of the perturbation wave number $k$~\citep{chandra81}. 
If the interface is sharp, the problem is mathematically ill-posed because the 
growth rate diverges with large $k$. The simulations we performed began with
completely sharp interfaces, but numerical diffusion in the simulation
regularizes the problem by increasing the width of the boundary
and thereby bounding the growth rate. In experiments, the growth rate is bounded 
either because surface tension limits the maximum 
$k$ or because physical diffusion widens the boundary~\citep{duff62,faber95}.

Our studies of the Rayleigh-Taylor problem consist of both single- and multi-mode 
simulations. The validation tests consisted of multi-mode simulations
performed from a standard set of initial conditions, allowing for comparison
with results of other research groups and experiments. 
Single-mode simulations allow for testing of convergence of 
the solutions and are meant to determine the minimum resolution per 
perturbation wavelength for a converged simulation. This case
is of particular interest because it illustrates the difficulty of code
validation even for apparently simple laboratory experiments.

In the case of a Rayleigh-Taylor configuration with a multi-mode 
perturbation, bubble and spike mergers and 
bubble/spike competition are thought to lead to an instability growth
according to a $t^2$ scaling law, which 
may be written as \citep{youngs94}
\begin{equation}  \label{eq:tsq}
h_{\mbox{\scriptsize{b,s}}} = \alpha_{\mbox{\scriptsize{b,s}}}gAt^2  \; ,
\end{equation}
where $h_{\mbox{\scriptsize{b,s}}}$ is the height of a bubble or spike, $g$ is the
acceleration due to gravity, and $A = (\rho_2 - \rho_1)/(\rho_2 + \rho_1)$ is
the Atwood number. $\rho_{1,2}$ is the density of the lighter 
(heavier) fluid, and $t$ is the time. The coefficient $\alpha$ 
is a measure of the rate of potential energy release. 
The initial conditions for the multi-mode simulations consisted of velocity 
perturbations corresponding to modes 32--64.
These initial conditions were adapted from a standard set
used by a consortium of researchers. This consortium, known as the 
Alpha Group, was formed in 1998 by Guy Dimonte for the purpose of 
investigating the validity of the $t^2$ scaling law, and, if it holds, to 
determine the value of $\alpha$.  
The study undertaken by the consortium includes both code-to-code comparisons 
and comparison with experiment \citep{dimonte01}.

Recent work on the Rayleigh-Taylor problem by \citet{cook01} and \citet{glimm01}
indicates a sensitivity of simulation results to initial conditions and
numerical diffusion effects. \citet{cook01} performed simulations of the
Rayleigh-Taylor instability for the case of incompressible, miscible fluids
with a 3:1 density ratio.
In this case, the solutions were evolved by the Navier-Stokes equations
augmented by a species transport-diffusion equation.
Cook and Dimotakis performed three large three-dimensional simulations
that differed only in the initial multi-mode perturbation, and their
results indicate a sensitive dependence of the growth rates on the initial
perturbation. In particular, they found that the growth rates for the
perturbation with the smallest characteristic wave number were the fastest,
and the growth rates for the perturbation with the largest characteristic
wave number were the slowest. In addition, they found that mixing had
an even greater dependence on the initial conditions, with their
smallest characteristic wave number simulation exhibiting the
``largest unmixedness."

\citet{glimm01} performed a study of the effects of numerical diffusion
on instability growth rates. The group performed three-dimensional simulations
with total variation diminishing (TVD) schemes and with a front tracking scheme.
The front tracking results differed from the TVD results by 40\%, with the TVD schemes
having the smaller rates. The group attributes this difference to the presence of
numerical diffusion in the TVD schemes. The TVD simulations were performed
with and without an artificial compression scheme to limit mass diffusion,
but showed similar results. From these similar results the group concludes that
the principal dissipative effect is viscosity, though they note that an alternate explanation is
that the artificial compression scheme does not sufficiently control the mass diffusion.
It should be mentioned that the initial conditions were not exactly the same
for the front tracking and TVD simulations, with the TVD simulations having
a lower mode number. Given the results obtained by \citet{cook01}, these differences
in initial conditions may play a role in the differences found between the TVD
and front tracking results, though the effect may be to decrease the differences.

\begin{figure}
\plotone{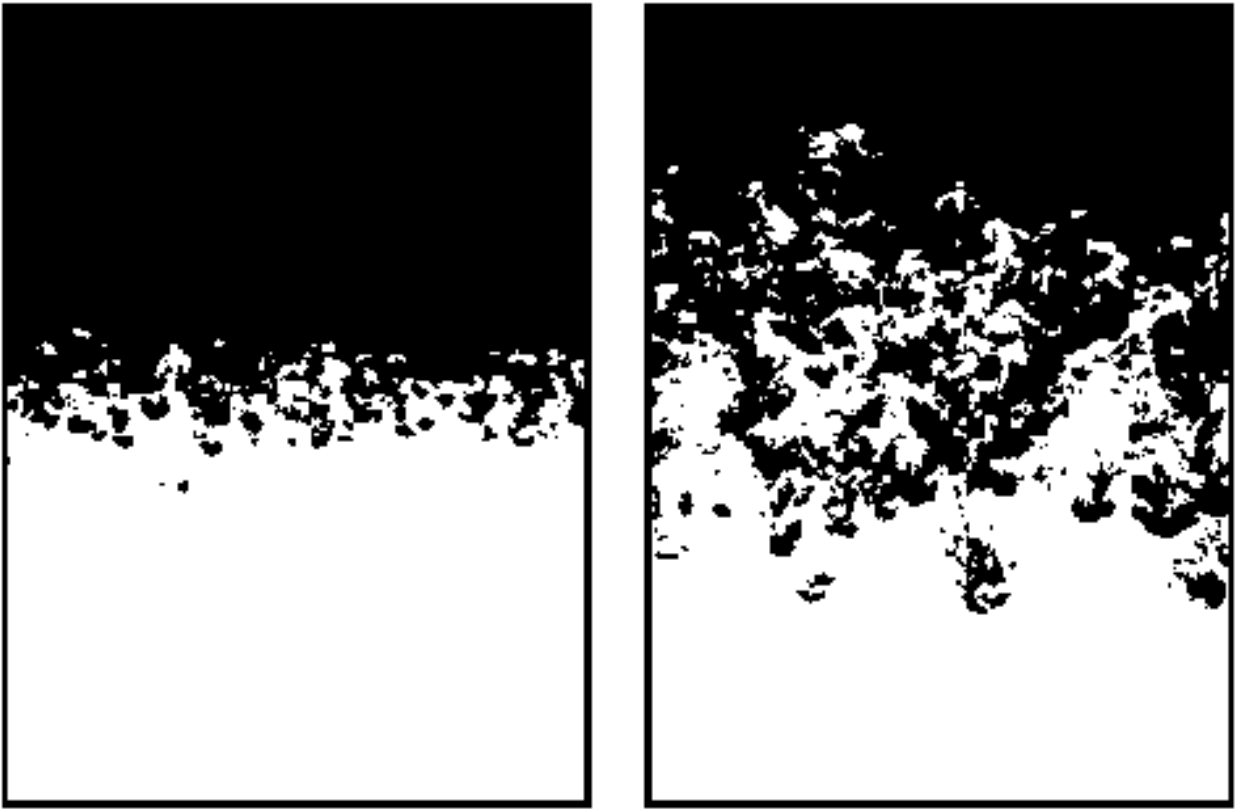}
\caption{Experimental results from a multi-mode Rayleigh-Taylor
experiment performed on the Linear Electric Motor. Shown are bi-level 
laser-induced fluorescence images from 
an experiment with Atwood number A = 0.32 at t = 25 ms (left) and 44 ms (right).
The dense material ($\rho_2 = 1.43\gcc$) is on the bottom and
appears white. The light material ($\rho_1 = 0.73\gcc$) is on
the top and appears black. The direction of the acceleration of the experimental
capsule was down, providing an effective upward acceleration.
The width of the material shown in each panel was 6.2 cm.
\label{fig:alpha_exp1}
}
\end{figure}

\begin{figure}
\plotone{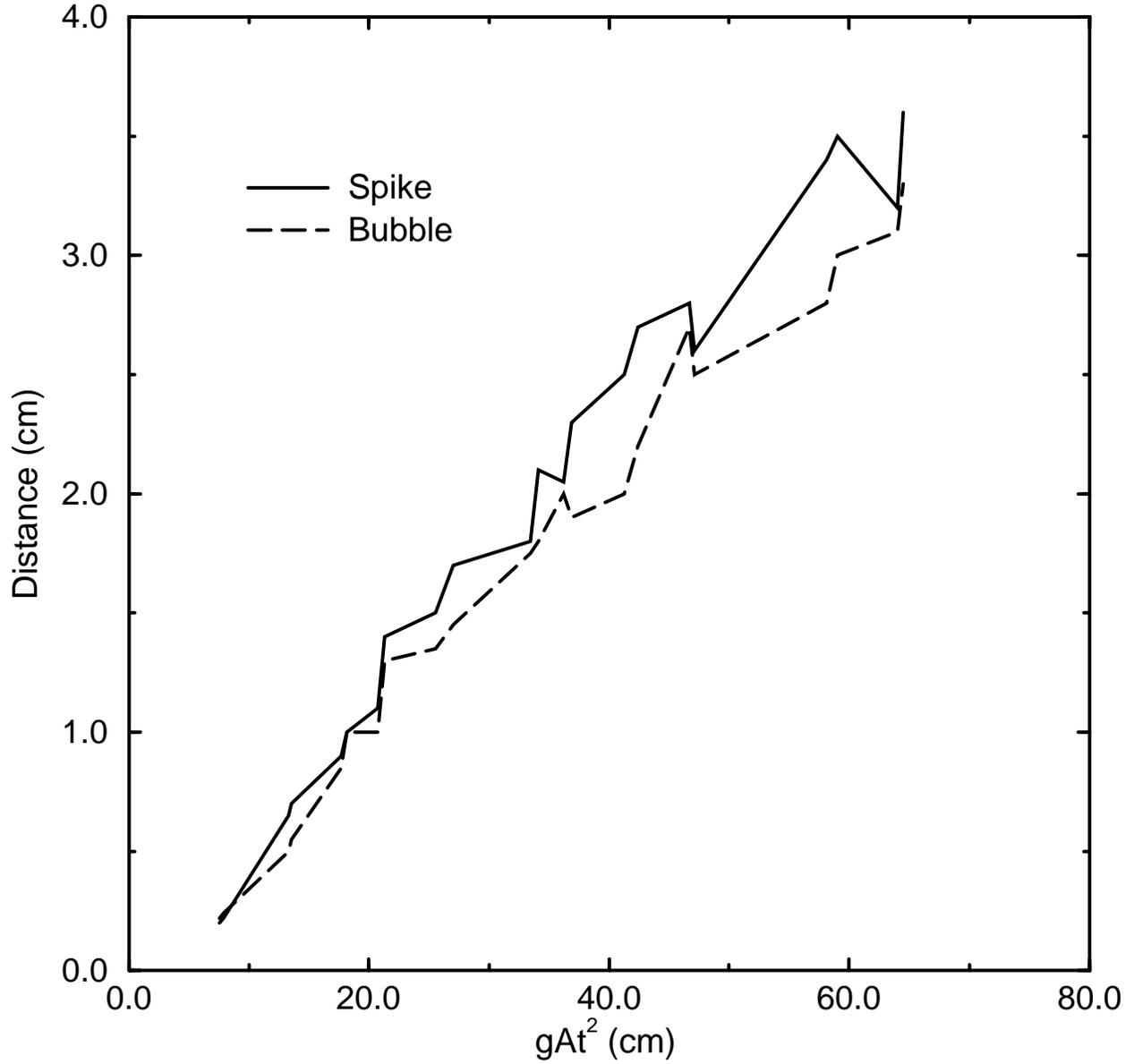}
\caption{Plot of distance vs.\  $gAt^2$ from a 
multi-mode Rayleigh-Taylor experiment performed on the Linear 
Electric Motor. Shown are
the magnitudes of bubble height and spike depth as functions of the product of
acceleration ($g$), Atwood number ($A$), and the time squared ($t^2$).
The slope of each curve equals $\alpha$, the rate coefficient. For
this experiment, fitting straight lines to the curves produced $\alpha = 
0.052 \mbox{ and } 0.058$ for the bubbles and spikes, respectively. 
\label{fig:alpha_exp2}}
\end{figure}

The experimental multi-mode Rayleigh-Taylor results come from studies performed by
\citet{schneider98} and \citet{dimonte00}. 
The experiments investigated the Rayleigh-Taylor instability over a range of 
density ratios using a variety of sustained and impulsive acceleration histories. 
The experiments consisted of a sealed plastic fluid container accelerated 
by the Linear Electric Motor \citep{dimonte96}. The
diagnostics used for comparison were provided by laser-induced
fluorescence, in which a dye is added to the 
system and excited by a laser beam focused
into a sheet propagating upwards through the cell. Images 
captured with a gated charged-coupled device and 35 mm film cameras
allow diagnosis of the internal structure of the
mixing zone. Figure \ref{fig:alpha_exp1} shows laser-induced fluorescence 
images from an experiment with $A = 0.32$ and a nominally constant acceleration
of $\sim 68$ times the acceleration of gravity on Earth. 
The images were captured at 25 and 44 ms.
The bi-level images shown 
are made from intensity images by setting the values to zero below and unity
above an intensity threshold. The threshold was chosen as $\sim$ 50\%
to conserve the two fluid volumes. In the images, the lighter material
is on the top and appears black, and the heavier material is on the bottom
and appears white.  Figure \ref{fig:alpha_exp2} shows the 
bubble and spike magnitudes
from the experiment. The resulting values for $\alpha$ were 0.052 for the
bubbles and 0.058 for the spikes.
Dimonte and Schneider report $\alpha \sim 0.05$ from several experiments
with a constant acceleration.  These results are in good  agreement with
earlier experimental work by \citet{youngs89}.

There are many unknowns that may influence the results of these
experiments. There may be unaccounted-for noise in the experiment that
can change the actual initial conditions.  Since the value of $\alpha$
is thought to be dependent on the power spectrum of the initial
perturbation, long wavelength noise in the experiment could make
$\alpha$ larger because these modes would dominate the dynamics in the
nonlinear regime. This situation will occur if the long wavelength
modes never reach the self-similar regime given the length and time
scales of the experiment.

In addition, the diagnostics of the experiments may lead to spurious
results. For example, because the laser-induced fluorescence method
illuminates the mixing zone with a planar sheet of light, this
diagnostic can lead to aliasing of long wavelength structures into
short wavelength features in the images, thus affecting the
interpretation of observed small-scale structures in the mixing zone.
Also, because of the dynamic limits on diagnostic resolution,
the formation of small-scale structure cannot be completely
determined.

\begin{figure}
\plotone{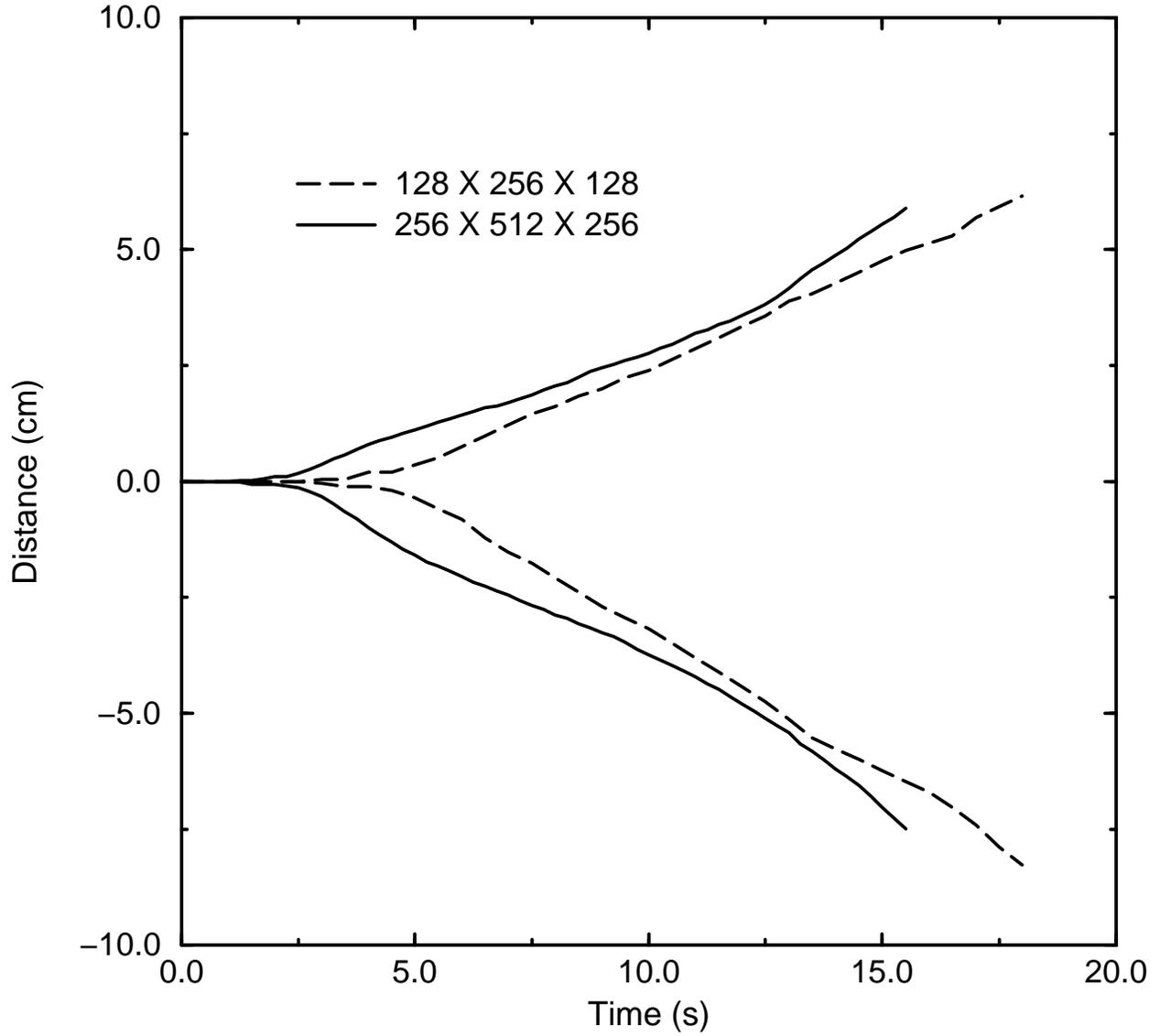}
\caption{Plot of distance vs.\ time from two three-dimensional multi-mode
simulations. 
Shown are bubble heights (the two top curves)  and spike depths (the two lower curves) 
as measured from the initial fluid interface.
The simulations had effective resolutions of $128 \times 256 \times 128$ and 
$256 \times 512 \times 256$. 
\label {fig:alpha_sim1}}
\end{figure}

\begin{figure}
\plotone{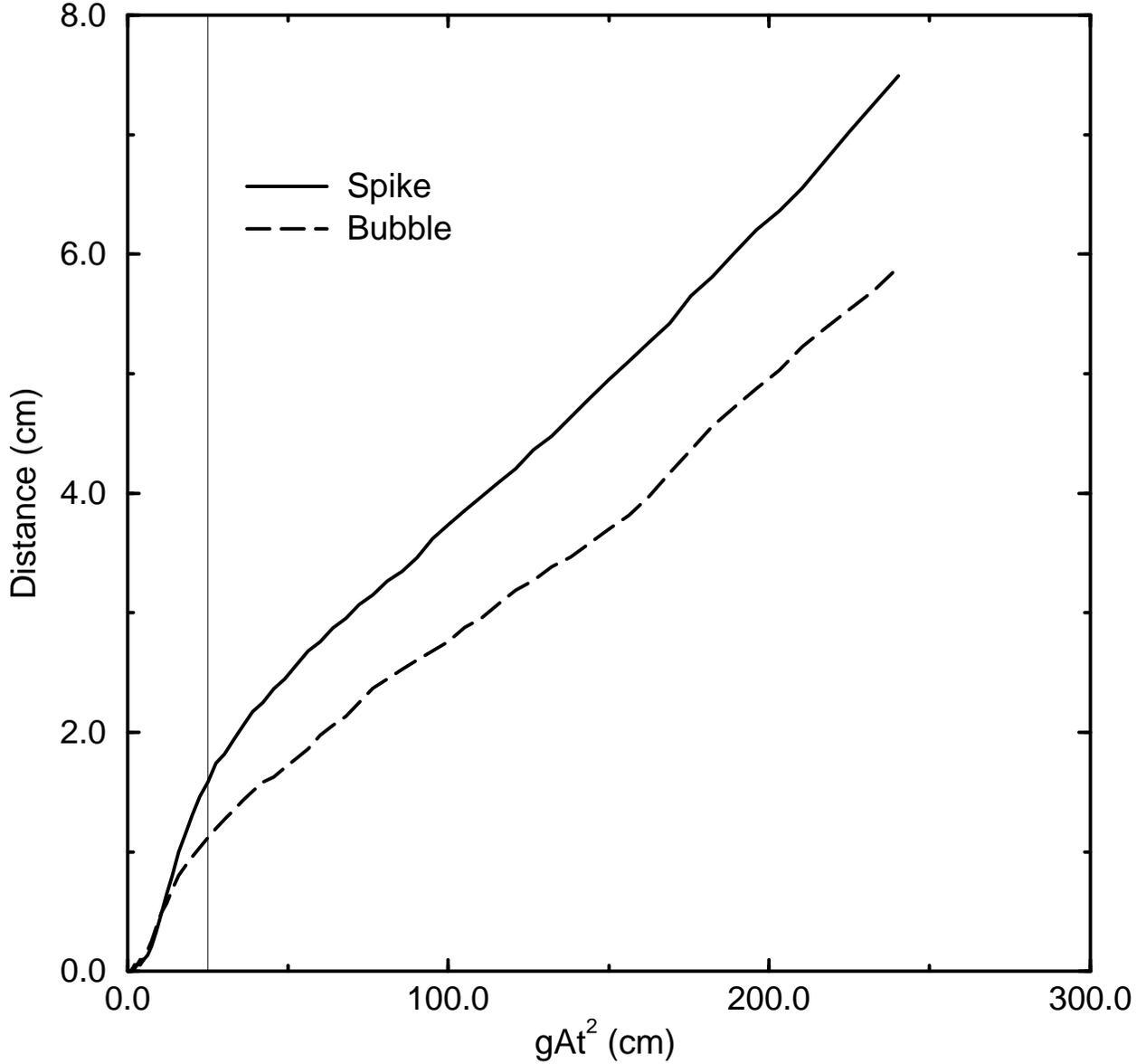}
\caption{Plot of distance vs.\  $gAt^2$ from the higher resolution
($256 \times 512 \times 256$) three-dimensional simulation. 
Shown are the magnitudes of bubble height and spike depth as functions of the 
product of acceleration ($g$), Atwood number ($A$), and the time squared ($t^2$).
The slope of each curve equals $\alpha$, the rate coefficient. 
For this simulation, fitting straight lines to the entire curves produced $\alpha = 0.024 
\mbox{ and } 0.030$ for the bubbles and spikes, respectively.
If the first five seconds of evolution are neglected to eliminate the parts of 
the curves with a rapidly changing slope, a straight line fit yields $\alpha = 0.021
\mbox{ and } 0.026$. The thin vertical line at $gAt^2 = 25$ marks $t = 5$ s.
\label {fig:alpha_sim2}}
\end{figure}

\begin{figure}
\epsscale{1.0}
\plotone{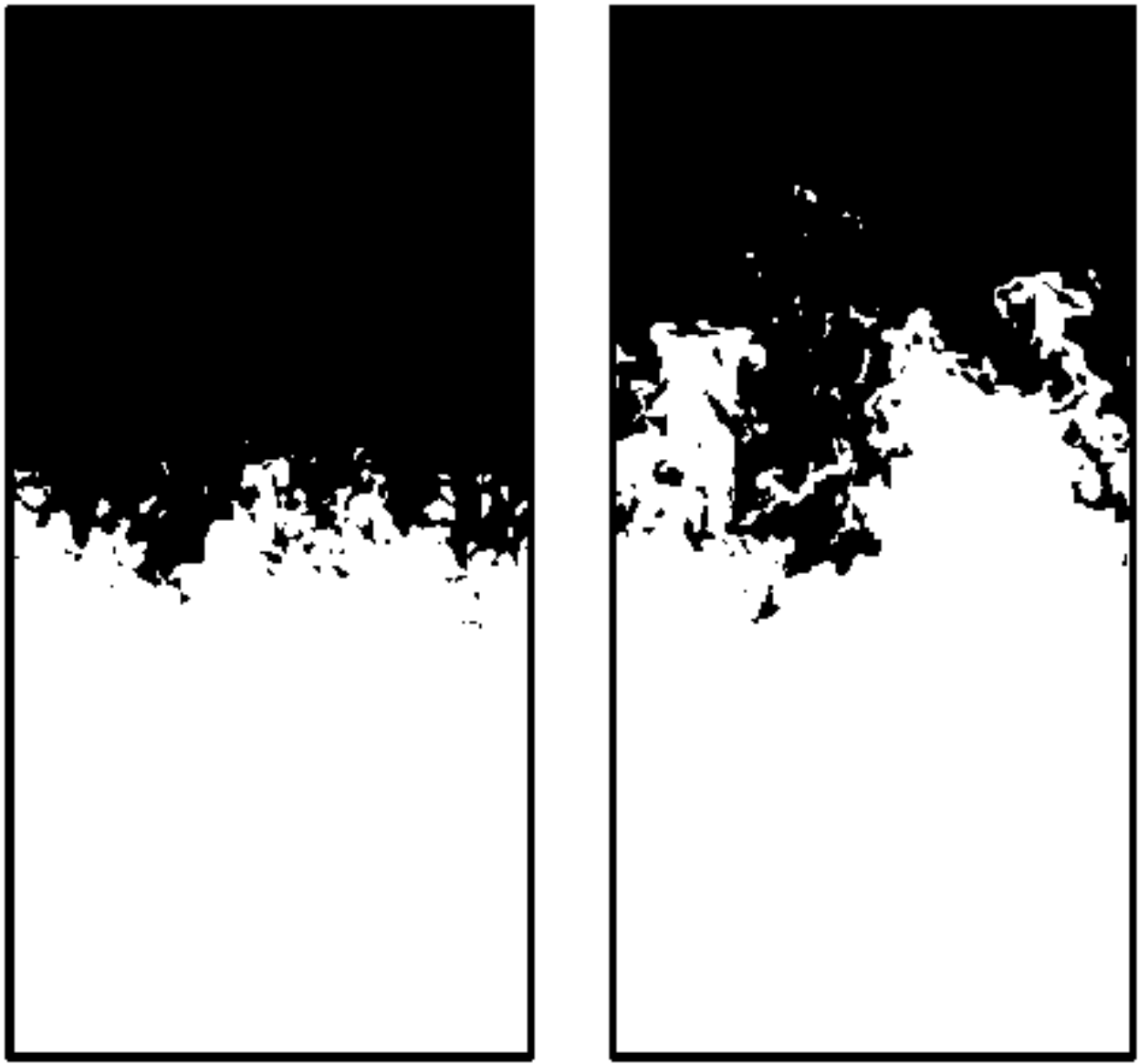}
\epsscale{1.0}
\caption{Bi-level cross sections of the multi-mode simulation at $t$ = 8.75 s (left)
and 15.5 s (right). The images were produced from cross sections of the 
heavy fluid abundance in the $x$-$y$ plane at $z$ = 2.5 cm. The images are
shown from the same perspective as those from the experiment with the
heavy fluid appearing white and the light fluid appearing black and an 
upward direction of the acceleration.
\label{fig:bi_sim}
}
\end{figure}

\begin{figure}
\epsscale{0.8}
\plotone{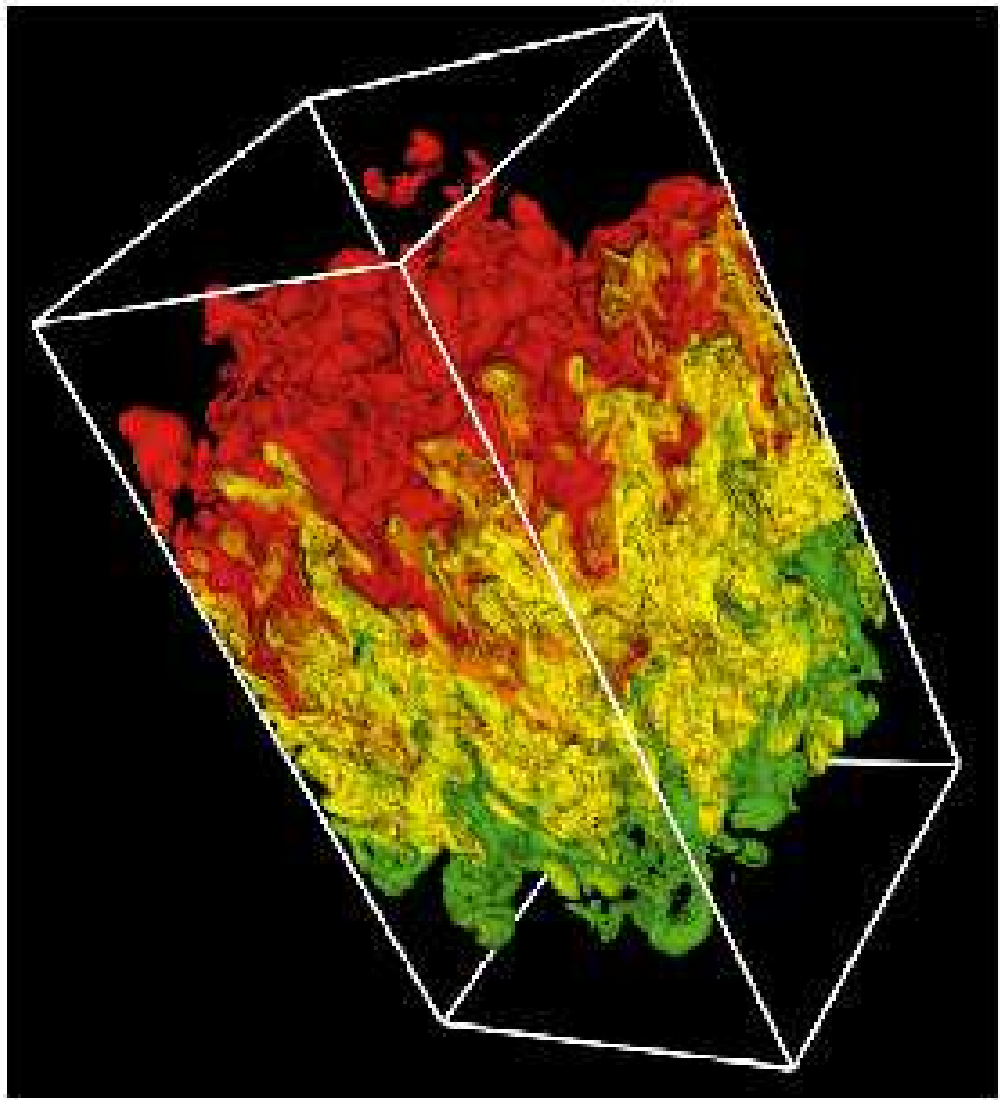}
\epsscale{1.0}
\caption{Rendering of the mixing zone of the higher resolution multi-mode
Rayleigh-Taylor simulation. Shown is density at a simulation time of 15.5 s. 
The colors indicate lower density (red), intermediate density (yellow), and
higher density (green). Densities higher or lower than those occurring
in the mixing zone are transparent. The initial perturbation consisted of
modes 32-64, with an effective resolution of $256 \times 512 \times 256$.
\label{fig:6levrt}
}
\end{figure}

Results of our multi-mode simulations are shown in 
Figures \ref{fig:alpha_sim1} - \ref{fig:6levrt}. The simulations began
from a Rayleigh-Taylor configuration with $A = \frac{1}{2}$ 
($\rho_1 = 1\gcc$ and 
$\rho_2 = 3\gcc$) and with $g = 2\cmss$. 
In the simulations, gravity acts in the $y$-direction, and the simulation domain 
consisted of a $10 \times 20 \times 10$ cm region, with the fluid interface 10.625 cm 
above the bottom of the domain. The initial interface displacement perturbation
was converted to a velocity perturbation using the linear theory of 
the Rayleigh-Taylor instability \citep{chandra81}. 
In order to completely resolve these initial perturbations, the mesh was forced
to be fully resolved in the perturbed region of the simulation domain. This
region was 4 cm tall and centered on the initial fluid interface.

Figure \ref{fig:alpha_sim1} shows bubble heights and spike depths 
for two three-dimensional multi-mode
simulations. The effective resolutions of the two simulations in the $(x,y,z)$ 
directions were $128 \times 256 \times 128$ and $256 \times 512 \times 256$.
The top two curves are bubble heights, and the lower two curves are spike depths.
The bubble heights and spike depths were measured by averaging the density
in each $x$-$z$ plane, and comparing the average density profile to the 
initial density profile, marking the points at which the average density
deviated from the initial profile by more than 1\%.
Results were robust to reasonable changes in this threshold value.
The distances shown were measured from the initial fluid interface, and
the distance between the two curves from each simulation is the width
of the mixing zone. The results show some differences between the two
simulations, but both show a predominantly linear growth above a certain 
point.

Figure \ref{fig:alpha_sim2} shows results from the higher resolution
simulation. Shown are bubble and spike magnitudes plotted vs.\  $gAt^2$ from the 
higher resolution three-dimensional simulation. 
The slope of a linear fit to each 
curve gives $\alpha$, the rate coefficient. 
The results for $\alpha$, obtained from a linear fit to each
curve, were  0.024 and  0.030 for the bubbles and spikes, respectively.
If we neglect the first five seconds as being part of a different
stage of evolution than the merger-dominated regime that gives the $t^2$ behavior
and look at the slopes, we find lower results for $\alpha$, 0.021 and 0.026.

Figure \ref{fig:bi_sim} allows for visual comparison of
the structure of the mixing zone of the higher resolution simulation to 
that of the experiment. Because the multi-mode simulations were not 
designed to match the experiments, (i.e. different Atwood numbers, 
accelerations, and geometry), we may make only a qualitative comparison.
Figure \ref{fig:bi_sim} shows bi-level images of the simulation at 
$t$ = 8.75 s. (left panel) and $t$ = 15.5 s. (right panel). The images
were created from the heavy fluid abundance in the $x$-$y$ plane at
$z$ = 2.5 cm by setting the values to zero below and unity above an 
abundance of 0.5. As with the experimental results, the dense fluid
is on the bottom and appears white, and the light fluid
is on the top and appears black. The early time was chosen to match
the proportion of evolution as that of the early time in the experimental
images. 
Figure \ref{fig:6levrt} is a rendering of density from the higher resolution 
simulation. Shown is the mixing zone, with well-developed bubbles and spikes.
The colors indicate lower density (red), intermediate density (yellow), and
higher density (green). The higher and lower density material above and
below the mixing zone is transparent.
As illustrated in these figures, the higher resolution 
multi-mode simulation 
shows a very similar structure to the experiments. 
Our results indicate $\alpha \leq 0.03$, however, which is not in good agreement
with the experiments and indicates the presence of some systematic error. 
Other research groups in the consortium using Eulerian
hydrodynamics methods report similar results, although groups using other 
methods report different results. Results of consortium studies 
will appear in publications of the Alpha Group \citep{dimonte01}.

Because our multi-mode simulations did not agree well with the
experimental results, we initiated a study of single-mode instabilities
to further investigate mesh resolution effects.
The single-mode simulations all began from initial conditions 
consisting of $A = \frac{1}{3}$ ($\rho_1 = 1\gcc$ and $\rho_2 = 2\gcc$) with 
$g = 1\cmss$. The initial perturbation 
consisted of a sinusoidal vertical velocity perturbation of 2.5\% of the local 
sound speed, with the horizontal components chosen to make the initial velocity field 
divergence-free, corresponding to the near-incompressible
nature of the experiments. The simulation domains were 0.25 cm $\times$ 1.5 cm for the 
two-dimensional simulations and 0.25 cm $\times$ 1.5 cm $\times$ 0.25 cm for the 
three-dimensional simulations. 
As with the multi-mode simulations, the grid was 
forced to be fully resolved in the region of the initial perturbation.

\begin{figure}
\plotone{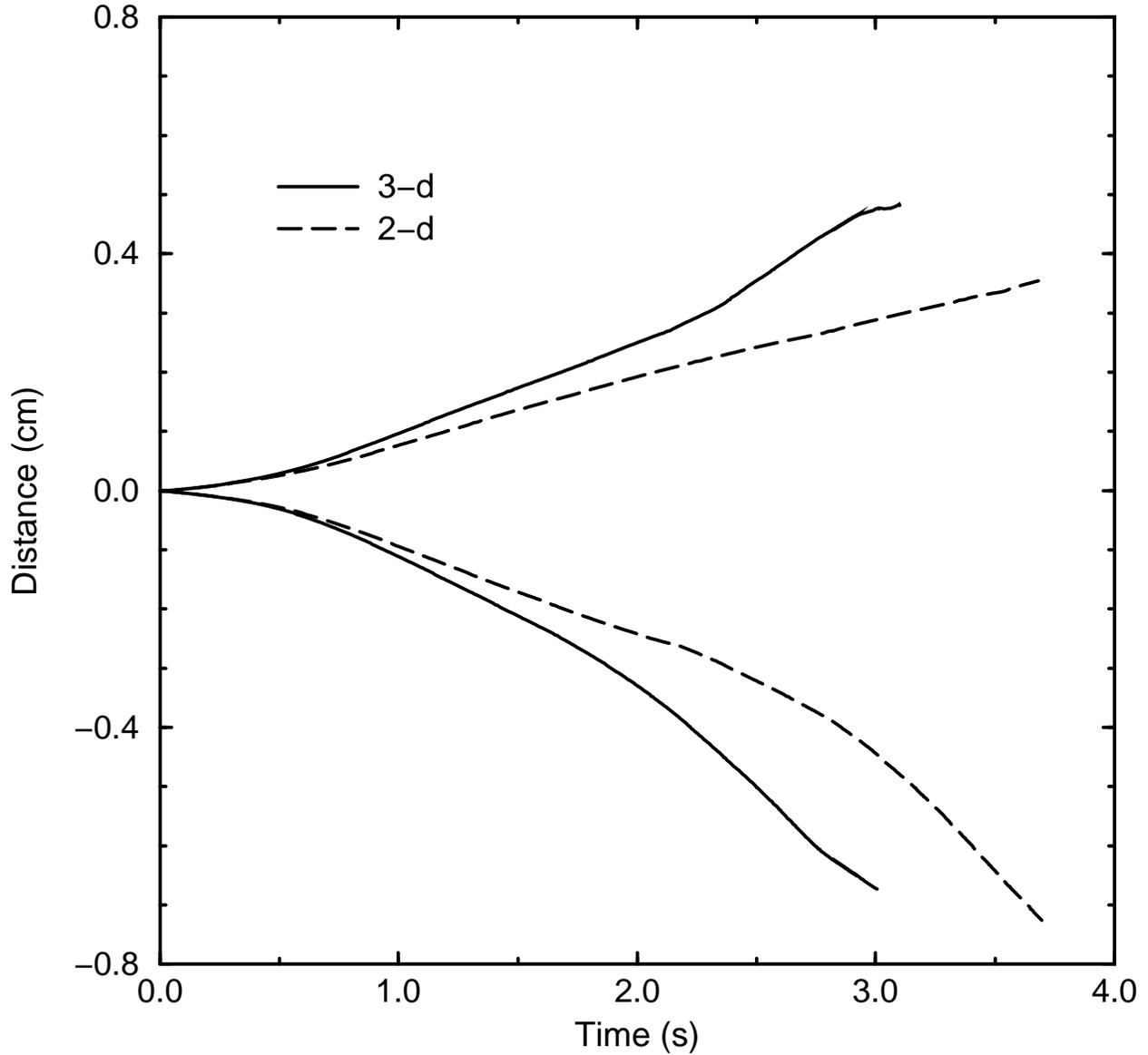}
\caption{Bubble heights and spike depths vs.\  time for two-dimensional and
three-dimensional simulations of single-mode Rayleigh-Taylor instabilities.
The top two curves are bubble heights, and the lower two curves are spike depths.
Each distance was measured from the initial fluid interface. The effective resolutions
were 128 $\times$ 256 (2-d) and 128 $\times$ 256 $\times$ 128 (3d).
\label{fig:single01}
}
\end{figure}

\begin{figure}
\epsscale{1.0}
\plotone{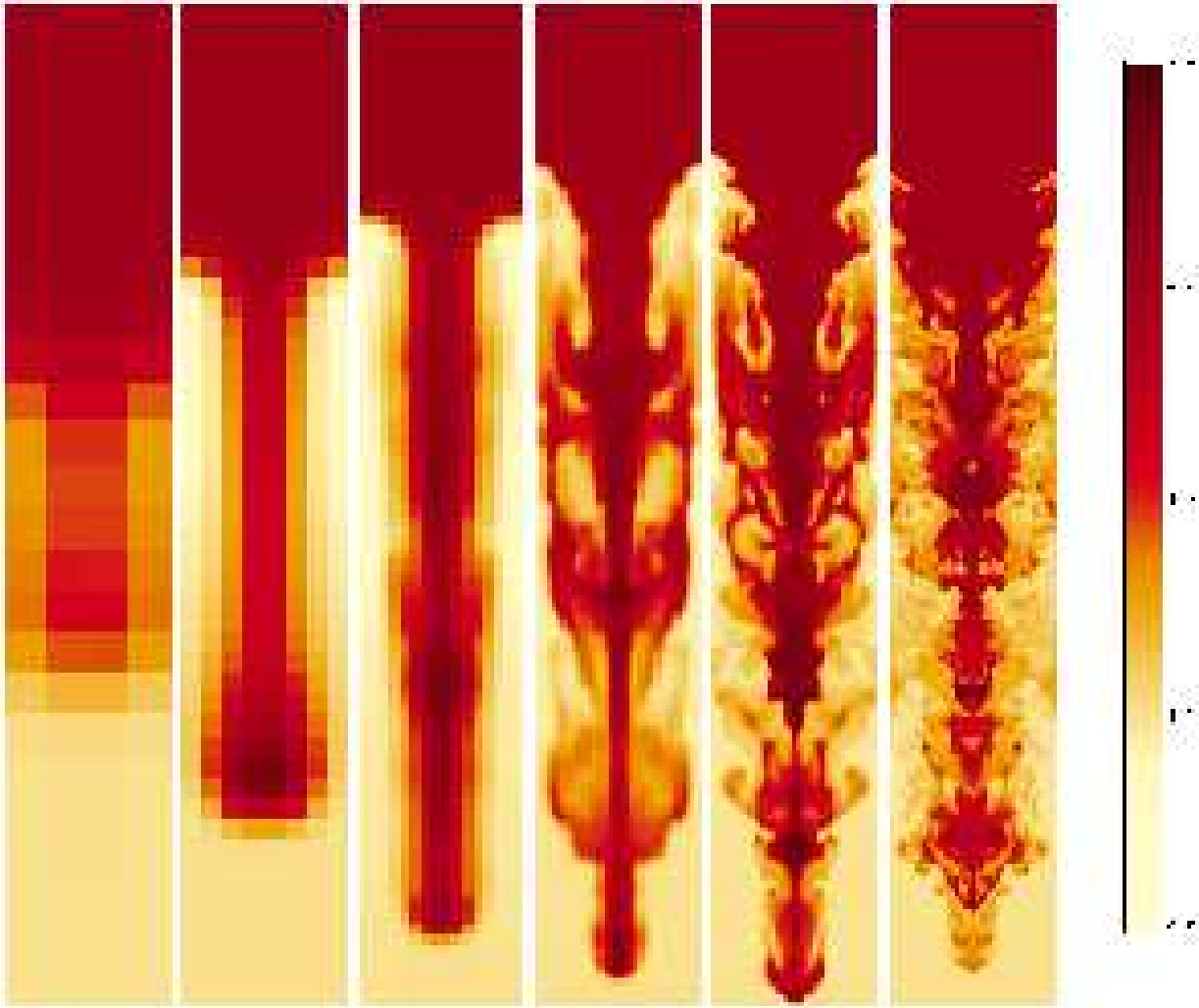}
\epsscale{1.0}
\caption{Plots of density (in a plane through the center of the domain)
from three-dimensional single-mode instability simulations at 3.1 s of evolution time. 
The effective resolutions for the panels  are, from left to right, 
$\lambda =$ 4, 8, 16, 32, 64, and 128 grid points. 
In the images, the value of each computational zone was determined from
the zone averaged values and no interpolation was done. The effect is that
one sees the zones as a series of squares of uniform color. This effect is
obvious in the panels from the lower resolution simulations.
\label{fig:single02}
}
\end{figure}

\begin{figure}
\plotone{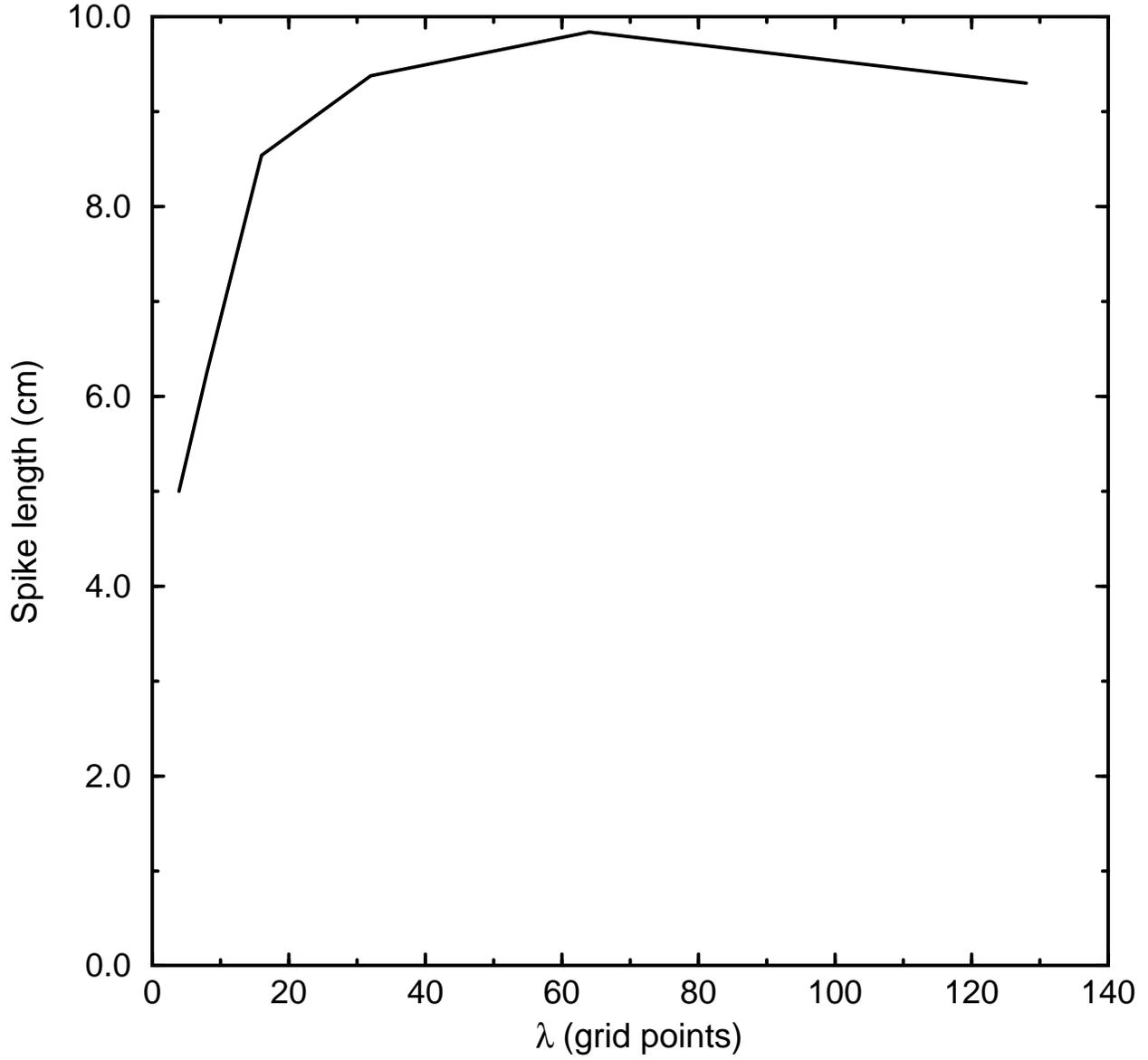}
\caption{Spike magnitude vs. resolution from the single-mode 
instability simulations. The points making the curve correspond to the spike 
magnitudes from the simulations for which density plots are presented in 
Figure~\ref{fig:spikes}.
\label{fig:spikes}
}
\end{figure}

Results from these simulations are shown in Figures \ref{fig:single01} - \ref{fig:spikes}. 
Figure \ref{fig:single01} shows plots of bubble height and spike depth for 
two simulations, one two-dimensional (dashed lines) and one three-dimensional (solid lines),
from equivalent initial conditions. The top two curves are bubble heights, 
and the lower two curves are spike depths.
The effective simulation resolutions are 128 $\times$ 768 (2-d) and 128 $\times$ 768 
$\times$ 128 (3-d). 
In these calculations, the bubble heights and spike depths were measured by tracking
the advection of each fluid and recording the positions of 99\% and 1\% abundances.
The distances shown were measured from the initial fluid interface. The distance between 
the two curves from each simulation represents the width
of the mixing zone. The results indicate that instability growth rates in 
three-dimensional simulations are larger than those found in the equivalent 
two-dimensional simulations as was 
observed by~\citet{kane00}~\citep[see also][for an incompressible comparison]{young01}. 

In addition, we find that obtaining a reasonably converged
estimate of the growth rate requires at least 25 grid points per 
wavelength of the perturbation, 
that grid noise seeds small-scale structure, and that the 
amount of small-scale structure increases with resolution. 
As noted with the results of the laser-driven shock experiments, 
this behavior is expected
because the dissipation mechanism in PPM operates on smaller and smaller
scales as the resolution is increased.
Another result is that poorly-resolved simulations exhibit a 
significant amount of mixing due to numerical diffusion.
Figure \ref{fig:single02} illustrates these results. The panels
are images of density after 3.1 s of evolution for six three-dimensional
simulations of increasing resolution from the same initial conditions. 
Shown in each panel is a cross-section of the simulation volume in the $y$-$z$ plane. 
In the simulations, the acceleration due to gravity acts in the $y$-direction. 
The effective resolutions are, from left to right, $\lambda$ = 4, 8, 16, 32, 64, and 
128 grid points. The mixing due to numerical diffusion is readily visible
in the lower resolution results, as is the trend toward increasing amounts
of small-scale structure with resolution. We note that our ``metric" for
measuring resolution in terms of grid point per wavelength assumes square or cubic 
mesh cells. Adding resolution along the direction of gravity without adding resolution
in directions perpendicular to gravity would decrease the amount of
mixing due to numerical diffusion without changing the number of grid points per
wavelength.

It is worth noting that the instability growth rate, an integral
property of the flow, does not converge with
increasing mesh resolution in the case of the single-mode Rayleigh-Taylor instability.
Figure \ref{fig:spikes} plots the spike length vs.\  wavelength of the perturbation 
($\lambda$) measured in number of grid points. The greatest spike length is not
found at the highest resolution.
We attribute the decrease in instability growth rate found in the highest resolution
to the increased amount of small-scale structure. Further studies of
the single-mode Rayleigh-Taylor and the role of small scale structure are underway,
and complete results will appear in \citet{calder01}. 

\subsection{A Note on Convergence}

As described above, our hydrodynamics model is inviscid.
Making this model solvable requires a dissipative mechanism,
but such a mechanism does not guarantee convergence.
In our case, because  the PPM dissipation mechanism 
does not include a 
fixed damping length scale, there can always be fluid motion on 
the smallest scale we
resolve. If those motions affect the bulk properties, then even
a measure of these bulk properties will not converge with increasing
mesh resolution. 
There exist differing
opinions in the literature regarding the effects of the PPM dissipation
mechanism on simulation results.

\citet{porter94}\citep[see also][]{porter00} investigated numerical diffusion 
of PPM in
two dimensions. They found that the numerical viscosity in 
PPM depends on velocity of the flow and the mesh
resolution. Recent numerical work by~\citet{sytine00}
performed convergence tests for PPM and Navier-Stokes
simulations of homogeneous compressible turbulence in three dimensions.
By studying kinetic energy, enstrophy, and energy power spectra, 
they found convergence of the PPM and Navier-Stokes solutions to the same limit.

Caution is warranted, however, about the future use of PPM for
simulating subsonic turbulent flows. \citet{garnier99}, in a study 
of decaying isotropic turbulence, found that the flow suffered from excessive
numerical damping on small length scales and that not all of the
properties of the turbulent flow were correctly reproduced.  This
dissipation makes shock-capturing schemes poor choices for use with
explicit subgrid-scale models.  Also, \citet{xu01} found that the
dissipation in Godunov methods depends on the character of the flow and
the mesh configuration, and is not always consistent with the
Navier-Stokes viscous terms.   Transonic turbulent flows, however, where
viscosity is not the primary dissipative mechanism, remains an open
research question.

Clearly, the limitations of present computing resources prevent us from
adequately resolving both the laser experiments and fully-developed
turbulent flows. The simulations we are capable of performing, however,
are approaching the resolution required to resolve these flows. Moore's
Law~\citep{moore65} would indicate that computing resources for these
types of flow may be available fairly soon. At that point, experiments
such as the laser-driven fluid dynamics experiments (particularly with
another generation of growth and improvements in diagnostics), could
serve as validation test beds for simulations of turbulent flows. 
Those astrophysical flows with dramatically larger Reynolds
numbers are not likely to be resolved in the foreseeable future.  The
lessons learned in modeling turbulent flows, though, should allow for the
improvement of hydrodynamic methods and the development of appropriate
subgrid models, making astrophysical simulations more realistic and
accurate. Further, if as we increase the resolution of our simulations
we eventually will converge with Navier-Stokes flows as indicated
by~\citet{sytine00}, we may be able to determine the correct amount of
small scale structure for these flows and thereby answer some of the
unanswered questions in this study.

\section{Conclusions}

In this paper, we presented the results of our efforts at
verification and validation of FLASH, our astrophysical simulation
code. We presented the results of new verification tests and the
results of two validation tests in which we were able to carefully
compare our simulation results to experimental results. These
tests served to increase our understanding of the physics relevant
to the problems of interest, to improve our simulation techniques,
and to build confidence in our results. 

\subsection{Verification}
The verification tests we performed for this study
indicate that the PPM hydrodynamics module is performing as expected.
Advection tests on a uniform mesh
showed slightly better than second order convergence of the
error with mesh resolution.  The PPM module, on a uniform mesh,
propagated a sound wave
for one period with approximately second order accuracy, and a shock wave, as
expected, to first order. In the case of a shock wave on an adaptive mesh,
the code showed the expected first-order convergence as well.
The code was able to hold a standing shock steady for several times steps, 
also, indicating the Riemann solver is working correctly.

The material equation of state, as well as the hydrodynamic solver,
must be well known and well behaved in order to obtain a meaningful
comparison with experiments.  Verification testing of a
test equation of state resulted in establishing a procedure for
testing consistency and quantifying the results. This procedure will
allow us to quickly test new equations of state as we develop
additional physics modules and address new astrophysical problems.

\subsection{Validation}

The results of simulating laser-driven shock experiments
show that we can capture the bulk properties of the flow, 
including the morphological properties of the resolvable 
structures. 
We observed the expected instabilities at the two material interfaces
that grew with time during the course of the simulations.
We performed a mesh resolution study that in general showed 
convergence of instability growth at the material interfaces. 
The resolution study did not completely demonstrate convergence, though, 
because the amount of small scale structure present in the simulations increased
with resolution as may be seen by a visual inspection of the 
results (Figure~\ref{fig:6ch}).
We attribute this increasing amount of small scale
structure to the fact that the amount of numerical dissipation
in our hydrodynamics method decreases with increasing resolution.
Visual inspection of the experimental results and our simulated radiographs
suggested that a simulation with six levels of adaptive mesh 
refinement better matched the observed small scale structure. We note, though,
that the correct amount of small scale structure in the experiments
is not likely captured in the experimental diagnostics because of resolution 
limits and because the radiograph produces a two-dimensional 
shadowgraph of a three-dimensional experiment, possibly averaging out 
small structures. 

Measurement of the lengths of the spikes allowed us to quantify the
results of the simulations, and we found that simulations with seven
and eight levels of adaptive mesh refinement agreed to within 5\% of
the highest resolution (nine levels) simulation at all times during the
evolution (Figure~\ref{fig:7andexp4}).   Complete convergence is
prohibited by the growth in small-scale structure with increasing
resolution.

The spike lengths at intermediate resolutions match the experimental
results very well, falling within the experimental errors.  We
interpret these as an important calibration result that indicates that
six and seven levels of adaptive mesh refinement are appropriate for
two-dimensional simulations of these particular experiments.  In
addition, the simulations showed that the shock propagating through the
corrugated material interface develops a perturbation that oscillates
and becomes planar at the second interface, as expected from theory. We
also observed the expected imprinting of the perturbation on this
second interface.

For this validation problem, comparison between simulation and
experiment is limited by the diagnostic resolution of the experiments
and the amount of physics included in the simulations.  Improvements to
experimental diagnostics will be made in the next generation of laser
experiments, and improvements to the simulations could be made by
inclusion of a more realistic material equation of state, adding the
walls of the shock tube, and modeling the laser-driven energy
deposition process.  Absent these improvements, we cannot
conclude that the good agreement of our results with the experiment
completely validates the simulations.

The results of simulations of a single-mode Rayleigh-Taylor instability
demonstrated the effect of resolution and dimensionality on growth rates.
The single-mode simulations established that $\sim$ 25 grid points per
mode are required for a reasonable estimate for the growth rate. This
calibration result suggests a minimum resolution for
simulations involving the Rayleigh-Taylor instability with our code.
This required resolution may make complete simulations of many problems
prohibitively expensive at the present time.  We confirm previous
findings \citep{kane00,young01} that instability growth rates in
three-dimensional simulations are larger than those found in the
equivalent two-dimensional simulations, which indicates limits on
results from two-dimensional models of astrophysical phenomena.  The
single-mode simulations also showed that, as with the laser-driven
shock simulations, one sees structure formation at smaller and smaller
scales as resolution is increased. The result is that increasing resolution will
not necessarily produce a converged flow. Further, the single-mode simulations
also showed that small scale structure in highly resolved simulations can 
have an effect on integral properties of the flow, in this case instability 
growth rates.

Our multi-mode Rayleigh-Taylor instability simulations, like the
simulations of the laser experiment, show that the code results agree
with the observed bulk properties of the flow: we observe a mixing zone
that is very similar in structure to that of the experiment.  Our simulations do
not, however, agree well with the experimental rate coefficient, $\alpha$;
our results are systematically lower.  This difference could be due to limitations
of either the simulations or the experiments. The experimental
initial conditions, which strongly effect the growth rate, are poorly
characterized.  Further, limitations on experimental diagnostics
may distort measured growth rates and mixing.

The simulations, on the other hand, may have been under-resolved.  In
our low-resolution simulation, most of the power in the initial
conditions was in modes which were resolved with only $2$--$4$ grid
points; in light of our single-mode results, this is far too low.
Increasing the resolution to resolve these modes with $4$--$8$ points
increases the observed growth rate, in (somewhat) closer agreement with
experimental results.

\subsection{Evaluation of V\&V Results}

The principal conclusions we draw from our efforts are that validation of 
an astrophysical simulation code is a difficult process and that
the process led us to unanticipated questions. The verification
tests we performed allowed for a careful quantitative study of the 
accuracy and convergence rates of FLASH. The exercise proved useful
in a variety of ways, not the least of which was developing techniques
for accurately comparing the results of our simulations
to accepted answers. The validation tests we
performed, while they did not conclusively validate the simulations, 
provided insight into the many issues involved in numerical modeling, and
served to calibrate the code and build confidence in our results.

V\&V testing allows for an assessment of error in the
code and progress in building confidence in the results. The process is
limited, though. Analytic solutions are typically available only for simple
problems, which may not serve as a strenuous tests of a particular code. Convergence
studies, though essential, do not imply that the converged answer is
correct. Code-to-code comparisons are another useful tool; they can be
more probing than analytic solutions, and can shed insight into the
behavior of different methods. But again, similar answers do not
imply correctness.

Comparing numerical and experimental results is also difficult.
Nonlinear systems typically have exponential sensitivity to initial
conditions, so that any unmodeled initial perturbation in the system can
greatly affect the results.  Experimental results may be limited, noisy
measurements of a poorly-characterized but real physical situations,
while the numerical results may be complete diagnostics of the situation
derived by a model of dubious applicability.  Comparisons must, therefore,
rely on choosing features which are obtainable from both results, yet
sensitive enough to the physics of interest that comparing the
features is a strong probe of the model and numerics.  The complexity
of this procedure means that doing it well requires feedback between
the computational and experimental researchers, with all parties
improving techniques and performing multiple runs.  The computational
simulations may be very expensive, however, and performing relevant
experiments is a costly and difficult process.   This makes it hard to
improve results and demonstrate reproducibility.

As our stated goal in this work was validating an astrophysical
simulation code, the assumptions in the models and their applicability
to both terrestrial and astrophysical flows warrants discussion.
Our numerical models did not have an explicit viscosity.
We saw, however, that the effects of numerical
dissipation cannot be avoided in simulations.  Other physical processes
that may play a role but are not included in the models include surface
tension, species diffusion, and thermal diffusion. With only Euler's
equations, it is impossible to adjust the relevant dimensionless
numbers (Schmidt, Reynolds, Rayleigh, and Prandtl numbers, for example)
to match the correct properties of the fluid.

Our use of models without an explicit viscosity resulted in an
increasing amount of small scale structure observed with
increasing resolution in simulations of both validation problems.
As mentioned above, limits on diagnostic
resolution prevent us from determining the correct amount of small scale structure
in both validation experiments.
Were the experimental diagnostics better, we could incorporate
physically-motivated terms for viscosity, surface tension, or diffusion into the
equations we evolve and calibrate their magnitudes until we obtain a converged 
flow with the correct amount of small scale structure. Proceeding in this manner 
(as well as including a physically-motivated equation of state)
would help us validate our simulations of these and similar terrestrial
experiments, but would be of little use in the astrophysical case, where
viscosity and surface tension play no dynamic role.

Even with better diagnostics these types of
experiments can only further validate the code if we model additional physics that are
not relevant to the astrophysical problems of interest. As we
mentioned above, flows within stars are expected to have a Reynolds number greater
than 10$^9$, for which the viscosity-free Euler equations are a better approximation, but
such flows are impossible to achieve either in the laboratory
or in current simulations.
Therefore, we  conclude that the terrestrial experiments we have simulated have
served to build confidence in the hydrodynamics module in FLASH, but that there
are certainly still limits on the strength of statements we can make about
the validity of FLASH simulations of laboratory experiments.

We end by repeating a point from our introduction, namely that verification and 
validation can determine only if a code returns an incorrect answer. 
Verification and validation cannot prove that a theoretical result, either numeric
or analytic, is correct. By doing a sufficient number of tests, however, one can 
significantly increase one's confidence in the results. Our efforts at validation, 
although they have presented many challenges and led to new questions 
such as the effect of small scale structure on bulk properties of
flows, have increased our confidence in the simulations produced by FLASH. 

\acknowledgements This work is supported in part by the U.S. Department of Energy
under Grant No. B341495 to the Center for Astrophysical Thermonuclear
Flashes at the University of Chicago, in part under the auspices of the 
U.S. Department of Energy by the University of California, Lawrence Livermore National 
Laboratory under contract No. W-7405-Eng-48, and in part by other U.S. Department
of energy grants.
K. Olson acknowledges partial support from NASA grant NAS5-28524, 
L. J. Dursi acknowledges support by the Krell Institute CSGF, and  
P. MacNeice acknowledges support from NASA grant NAS5-6029. 
The work of T. Plewa was partly supported by the grant 2.P03D.014.19 from
the Polish Committee for Scientific Research. The authors thank Ewald M\"uller
for providing the electron-positron equation of state used in the study.
The authors also thank Robert Kirby for helpful comments.
The authors thank Mike Papka and the Argonne National Laboratory for 
visualization support and Ed Brown, Jim Truran, 
and Margaret Pepperdene for previewing 
this manuscript. Finally, the authors thank the anonymous
referee for his insightful criticism that greatly improved this work.

\clearpage

\clearpage

\clearpage

\clearpage

\clearpage

\clearpage

\clearpage

\clearpage

\clearpage

\clearpage

\clearpage

\clearpage

\clearpage

\clearpage

\clearpage

\clearpage

\clearpage

\clearpage

\clearpage

\clearpage

\clearpage

\clearpage

\clearpage

\clearpage

\clearpage

\clearpage

\clearpage

\begin{table}
\begin{center}
\caption{
Percentages of the failures at several required accuracies for the 
the electron/positron equation of state and 5 Riemann solver iterations.
\label{tbl:1}}
\begin{tabular}{|c|c|}  \tableline
    required accuracy   &   cases failed [\%] \\ \hline
        $10^{-1}$       &    0.02 \\
        $10^{-2}$       &    6.7 \\
        $10^{-3}$       &   21 \\
        $10^{-4}$       &   34 \\
        $10^{-5}$       &   43 \\
        $10^{-6}$       &   49 \\ \hline
\end{tabular}
\end{center}
\end{table}

\end{document}